\documentclass[%
 aip,
 amsmath,amssymb,
 reprint,%
]{revtex4-1}

\usepackage{graphicx}
\usepackage{dcolumn}
\usepackage{bm}

\usepackage[utf8]{inputenc}
\usepackage[T1]{fontenc}
\usepackage{mathptmx}
\usepackage{etoolbox}
\usepackage{xcolor}
\usepackage{comment}

\newcommand\thbn{\vartheta_{\rm Bn}}

\makeatletter
\def\@email#1#2{%
 \endgroup
 \patchcmd{\titleblock@produce}
  {\frontmatter@RRAPformat}
  {\frontmatter@RRAPformat{\produce@RRAP{*#1\href{mailto:#2}{#2}}}\frontmatter@RRAPformat}
  {}{}
}%
\makeatother
\begin{document}

\preprint{AIP/123-QED}

\title[Criteria for ion acceleration in laboratory magnetized quasi-perpendicular collisionless shocks]{Criteria for ion acceleration in laboratory magnetized quasi-perpendicular collisionless shocks: when are 2D simulations enough?}
\author{Luca Orusa}
\affiliation{Department of Astrophysical Sciences, Princeton University, Princeton, New Jersey 08544, USA}
\affiliation{Department of Physics and Columbia Astrophysics Laboratory, Columbia University, New York City, New York 10027, USA}
\author{Vicente Valenzuela-Villaseca}
\affiliation{Department of Astrophysical Sciences, Princeton University, Princeton, New Jersey 08544, USA}

\email{luca.orusa@columbia.edu; valenzuelavi1@llnl.gov}

\date{\today}

\begin{abstract}
The study of collisionless shocks and their role in cosmic ray acceleration has gained importance through observations and simulations, driving interest in reproducing these conditions in laboratory experiments using high-power lasers. In this work, we examine the role of three-dimensional (3D) effects in ion acceleration in quasi-perpendicular shocks under laboratory-relevant conditions. Using hybrid particle-in-cell simulations (kinetic ions and fluid electrons), we explore how the Alfvénic and sonic Mach numbers, along with plasma beta, influence ion energization, unlocked only in 3D, and establish scaling criteria for when conducting 3D simulations is necessary. Our results show that efficient ion acceleration requires Alfvénic Mach numbers $\geq 25$ and sonic Mach numbers $\geq 13$, with plasma-$\beta \leq 5$. We theoretically found that, while 2D simulations suffice for current laboratory-accessible shock conditions, 3D effects become crucial for shock velocities exceeding 1000 km/s and experiments sustaining the shock for at least 10 ns. We surveyed previous laboratory experiments on collisionless shocks and found that 3D effects are unimportant under those conditions, implying that 1D and 2D simulations should be enough to model the accelerated ion spectra. However, we do find that the same experiments are realistically close to accessing the regime relevant to 3D effects, an exciting prospect for future laboratory efforts. We propose modifications to past experimental configurations to optimize and control 3D effects on ion acceleration. These proposed experiments could be used to benchmark plasma astrophysics kinetic codes and/or employed as controllable sources of energetic particles.
\end{abstract}

\maketitle

\section{Introduction}
\label{sec:intro}

Non-relativistic, magnetized collisionless shocks are ubiquitous structures in the universe. These systems are characterized by having the ion-ion mean free paths that far exceed the density gradient length-scale associated with the shock discontinuity. Therefore, energy and momentum transfer are not mediated by Coulomb binary collisions between particles but rather through collective electromagnetic interactions. Examples of collisionless shocks in astrophysics are supernova remnants (SNRs), planetary bow-shocks, and galaxy cluster shock waves. Additionally, collisionless shocks are widely regarded as efficient sites for particle acceleration, playing a crucial role in the production of cosmic rays (CRs) \cite{axford+77p, bell78a, blandford+78, morlino+12, caprioli12}. 

The conditions governing particle energization in these shocks are determined by a relatively small set of key parameters: the Alfv\'enic Mach number ($M_A = v_{\text{sh}}/v_A$, where $v_{\text{sh}}$ is the shock velocity and $v_A = B_0/\sqrt{\mu_0\rho}$ is the Alfv\'en velocity, $B_0$ the upstream magnetic field, $\mu_0$ the permittivity of vacuum, and $\rho$ the plasma mass density), the thermal plasma-$\beta$ parameter (the ratio of thermal to magnetic pressure, $\beta = p_{th}/p_{M}$), and the angle $\thbn$ between the shock propagation direction and the upstream magnetic field $B_0$. 

In this work, we focus on the so-called high-$M_A$ regime ($M_A>15$) of \emph{quasi-perpendicular} shocks ($\thbn > 60^\circ$), which is relevant to several astrophysical environments. For example, the quasi-perpendicular region of the Earth's bow-shock, where $M_A \lesssim 20$ and $\beta \sim 1$, is known to efficiently accelerate ions \cite{johlander+21, Lalti+22, wilson+16b}. Similarly, SNRs are also widely associated with cosmic ray acceleration. A particularly interesting case is SN 1006, where the local magnetic field direction $B_0$ has been determined \cite{rothenflug+04, bocchino+11, gamil+08, giuffrida+22, SN1006HESS}. The remnant exhibits an azimuthally symmetric radio emission pattern \cite{rothenflug+04}, suggesting efficient particle acceleration at least at GeV energies across parallel, oblique, and perpendicular regions. Additionally, young extra-galactic supernovae associated with radio emissions may also feature quasi-perpendicular shock geometries \cite{chevalier+06}. On the largest scales of the universe, collisionless shocks are formed when galaxy clusters collide and merge. Observations of radio relics provide strong evidence for the acceleration of relativistic electrons at these merger shocks \cite{brunetti+14, willson70, fujita+01, govoni+04, vanweeren+10, lindner+14}. These shocks, typically characterized by a quasi-perpendicular configuration, propagate through the hot intracluster medium (ICM), a diffuse, weakly magnetized plasma with high temperature and a high plasma beta ($\beta \gg 1$).

Significant numerical efforts have been dedicated to studying perpendicular shocks, exploring their parameter space and the role of dimensionality in simulations. Particle-in-cell (PIC) simulations of low-$\beta$ quasi-perpendicular shocks have been conducted in 1D \cite[e.g.,][]{shimada+00,kumar+21,xu+20}, 2D \cite[e.g.,][]{amano+09a,bohdan+21,kato+10,matsumoto+15}, and small-box 3D setups \cite[e.g.,][]{matsumoto+17}, yet compelling evidence of particle acceleration remains elusive. A key finding from these studies is that in the quasi-perpendicular regime, the ion spectrum remains unchanged between 1D and 2D simulations, showing no evidence of non-thermal tails. Henceforth, we will discuss discrepancies between 2D and 3D simulations bearing in mind that the same differences exist between 1D and 3D. 

Recently, the strong constraints on the magnetic field orientation and ion acceleration were relaxed via more general simulations in three dimensions. Orusa \& Caprioli \cite{orusa+23} conducted an extensive campaign of hybrid particle-in-cell simulations (kinetic ions and fluid electrons) of low-$\beta$ quasi-perpendicular shocks with $M \gtrsim 25$, demonstrating for the first time in self-consistent kinetic simulations that a significant non-thermal ion population emerges only in 3D. This result contrasts with lower dimensionality  (1D and 2D) PIC and hybrid simulations of quasi-perpendicular shocks, where efficient ion acceleration remains challenging  \cite{amano+07, xu+20, guo+14a, kato+10, kumar+21, guo+14b, morris+23, ha+21, ha+22, bohdan+19a, bohdan+21, amano+22, matsumoto+15, matsumoto+17, kucharek+91, giacalone+93, giacalone+97, giacalone05, lembege+04, caprioli+14b, caprioli+14c, caprioli+15, caprioli+17, caprioli+18, haggerty+20, caprioli+20}.

They showed that in 2D simulations, particles are typically advected into the downstream region after at most one gyro-motion, preventing them from repeatedly crossing the shock and returning to the upstream. This is because, in 2D, magnetic field lines effectively act as walls, significantly reducing the probability of return. In contrast, a fully 3D magnetic field structure introduces enough degrees of freedom  for particles to leak back into the upstream\cite{jones+98} through three-dimensional trajectories and gain energy with each cycle via shock drift acceleration (SDA). As a result, 3D effects play a crucial role in accurately capturing shock dynamics and particle energization, which are often underestimated in 2D simulations. 

In general, the energy spectrum can be modeled as a power law $\propto E^{-\alpha}$. In Orusa \& Caprioli \cite{orusa+23} they found that that the higher is $M_A$, the "harder" the energy spectrum, that approaches $\alpha \sim 1.5$ for high-$M_A \gtrsim 100$ (corresponding to $\propto p^{-4}$ for non-relativistic particles), consistent with the universal spectral slope expected at strong shocks. For lower values of $M_A$, the spectrum becomes steeper, with non-thermal tails that progressively shrink and disappear for $M_A < 10$, showing no detectable difference from the 2D case in this low-$M_A$ regime.
A key factor in particle injection is the post-shock magnetic turbulence, which grows\cite{kato+10,bohdan+21,matsumoto+15} with $\propto \sqrt{M_A}$ . Higher levels of turbulence enhance the probability of ions returning upstream, leading to harder spectra.

A different regime describes the more weakly-magnetized astrophysical environments, such as galaxy clusters, that host high-$\beta$ oblique shocks. This class of shock has been investigated using both 2D PIC simulations \cite{guo+14a, guo+14b, xu+20, ha+21, ha+22, ha+23, kang+19} and 2D-3D hybrid simulations \cite{boula+24}, showing a preference for electron rather than ion injection \cite{xu+20}. Moreover, differences between 2D and 3D hybrid simulations appear to be minimal \cite{boula+24}, as neither exhibit non-thermal ion populations, though definitive conclusions have yet to be reached.

These exciting discoveries on collisionless shock astrophysics has sparked the interest of the experimental plasma physics community, who seek to reproduce astrophysics-relevant shock conditions and test astrophysical theories using laboratory experiments (see e.g.\cite{Takabe2021}).  Much of the progress on Earth-based experiments has been done using high-power, high-energy laser systems since they can create hypersonic pistons that propagate through an upstream medium, creating a shock at sufficiently high speeds so that the ion-ion mean free path far exceeds the system size. The interplay between astrophysics and laboratory plasma physics offers a unique and stimulating opportunity to test and constrain models of collisionless shock formations, plasma instabilities, and particle acceleration in controlled conditions.

In the past decade, the first successful generation of collisionless shocks in laboratory laser-driven plasma experiments \cite{Schaeffer2012,schaeffer+19,fiuza+20,yamazaki+22, yao+21} has been achieving conditions characterized by values of $M_A$ and $M_s$ relevant to the heliosphere and other astrophysical environments. Typically, these experiments have generated shocks lasting a few nanoseconds, corresponding to several ion gyro-periods. As noted by Orusa \& Caprioli \cite{orusa+23}, the acceleration process in quasi-perpendicular shocks is extremely fast (of the order of ten ion gyro-periods) and could potentially be tested in the laboratory. Quasi-perpendicular shocks form quickly as the magnetic field directly opposes the incoming plasma flow, enhancing compression efficiency. This is in contrast to \emph{quasi-parallel} shocks (where $\vartheta_{Bn} \leq 60^\circ$) that develop more gradually, as the magnetic field is aligned with the direction of shock propagation, leading to a slower shock formation mediated by multi-scale plasma processes \cite{caprioli+15,hada+03}. As a result, perpendicular shocks are easier to generate in the laboratory, where the available laser drive duration limits the overall experimental time-frame. Nevertheless, laboratory experiments have already found evidence of particle energization\cite{schaeffer+19, fiuza+20, yamazaki+22, yao+21} in the moderate to high Alfv\'enic Mach number regime (i.e., $M_A$ ranging between 4 to 30).

The pioneering experiments conducted by Schaeffer et al. \cite{Schaeffer2012,schaeffer+19} at the Large Plasma Device (LAPD) and the OMEGA laser facility \cite{boehly+95} marked the first laboratory observations of time-resolved electron and ion velocity distributions in magnetized perpendicular collisionless shock precursors. Yamazaki et al. \cite{yamazaki+22} investigated the formation of quasi-perpendicular supercritical magnetized collisionless shocks using the Gekko-XII HIPER laser system, while Yao et al. \cite{yao+21} conducted an experiment at the LULI2000 facility, where a laser-driven piston was used to generate an expanding plasma that propagated into an ambient hydrogen plasma within a uniform external magnetic field, producing a collisionless shock. We will explore these experiments in more detail below. It is also worth mentioning that Weibel-mediated collisionless shocks, have been successfully created at the National Ignition Facility, yielding new valuable insights about electron acceleration in turbulent shocks \cite{fiuza+20}.

This paper builds on the work of Orusa \& Caprioli \cite{orusa+23} by asking the question: are the existing 1D and 2D particle-in-cell simulations enough to model ion acceleration in these experiments or should 3D effects be considered? To do so, we extend the analysis of the parameter space, focusing on the conditions relevant to laser-driven laboratory experiments. Moreover, we introduce new scaling criteria exploiting our numerical results. We conducted a parametric study using a new set of simulations, focusing on the first tens of ion cyclotron times and examining the shock structure and accelerated ions in conditions with $M_s$ and $M_A$ in the range 5 to 30. We find that 2D simulations are adequate for all the experiments we surveyed. However, we predict that if one were to drive shocks $50\%$ faster that these experiments, then 3D modeling would be necessary to accurately calculate the energized ion spectra. On the basis of our findings, we propose a set of experimental configurations that could maximize ion acceleration, guiding future laboratory campaigns toward conditions where perpendicular shocks can efficiently energize particles.

There is an important distinction between our simulations and laboratory experiments. Typically, in the laboratory, a laser heats up a solid, launching a piston that expands into an ambient upstream plasma.  This piston compresses the upstream ambient magnetic field through the coupling of the lightest ion species in the piston and the upstream ions \cite{Schaeffer2017,schaeffer+19,yao+21}. A compression wave forms quickly, within 1–2 ion cyclotron times, and as the shock develops, it detaches from the piston, creating a downstream region behind the resulting density jump. Once the piston and shock are decoupled, the shock is sustained between the uncompressed upstream ambient ions and the ambient ions that have been swept into the downstream region. In contrast, our simulations do not launch a piston-driven plasma into an upstream medium, but they begin with a supersonic flow propagating toward a reflecting wall. The interaction between the incoming and reflected ion streams compresses the magnetic field and increases the density, leading to shock formation. Therefore our simulations are agnostic to what a realistic piston would do in an experiment, but this is a valid assumption on timescales after the shock-piston have decoupled, which is the case in our analysis. In other words, since the ion acceleration occurs only after the shock is fully formed and a downstream exists, neglecting the piston is a valid approximation over which energetic ions are produced and detected.

We emphasize that the simulations presented here are not intended to be accurate models of laboratory experiments, nor are they intended to fully replicate laboratory setup. To do so, one would need to resolve both the electron and ion dynamics, calculate the laser deposition on a solid-density target, ionization processes, coupling of specific ion species with the upstream medium, and other complications which would make the simulations computationally prohibitively expensive. Rather, we seek to offer theoretical guidance (numerical and analytical) for assessing the need of accounting for 3D effects to model particle acceleration in conditions relevant to laboratory experiments.

The paper is organized as follows: in Section \ref{sec:methods}, we present the details of the simulations performed. In Section \ref{sec:results}, we outline the simulation results. Section \ref{sec:scaling} discusses the parameter space relevant to laboratory experiments and provides the scaling equation for identifying optimal experimental setups. In Section \ref{sec:discussion}, we review previous experiments and propose new configurations that could exhibit strong ion acceleration. Finally, in Section \ref{sec:conclusions}, we summarize our conclusions.

\section{Methods}
\label{sec:methods}

All results presented in this work are obtained from simulations performed using the hybrid particle-in-cell {\tt dHybridR} code \cite{haggerty+19a} (kinetic ions and fluid electrons) in the non-relativistic regime \cite{gargate+07}. As explained in the introduction, our simulations do not model the piston but focus solely on the upstream ambient plasma; accordingly, all quantities reported below refer to the upstream plasma. In the simulations, a supersonic flow with speed $v_{sh}$, propagates towards a reflecting wall (left boundary). The interaction between the incoming and reflected ion streams generates a shock that moves rightward (along the $x$-axis), into a static and homogeneous perpendicular $B_0$ field with $\thbn=90$° along the $y$-axis. As a result, the downstream region remains stationary, and the kinetic energy of the incoming flow is efficiently transformed into thermal energy at the shock front.

Lengths are expressed in units of the ion skin depth $d_i \equiv c/\omega_p$, where $c$ is the speed of light and $\omega_p \equiv \sqrt{Z^2e^2n / \epsilon_0 m}$ is the ion plasma frequency, with $m$, $Z$, $e$, $n$, $\epsilon_0$ are the ion mass, charge state, fundamental charge, number density, and permittivity of free space, respectively. Time is measured in units of the inverse ion cyclotron time $\omega_c^{-1} \equiv m/(eB_0)$. Velocities are normalized to the Alfvén velocity $v_A \equiv B_0/\sqrt{\mu_0 m n}$ ($\mu_0$ is the magnetic permeability of vacuum), and energies to the kinetic energy per ion co-moving with the shock, $E_{sh} \equiv m v_{sh}^2/2$. The simulations include all three spatial components of the particle momentum and the electromagnetic fields. The hybrid model requires an explicit choice for the electron equation of state, and in this work, electrons are treated as adiabatic with an index $\gamma = 5/3$ \cite{caprioli+14a,haggerty+20,boula+24}. The choice of $\gamma$ in our simulations is a prescribed assumption, as we do not employ tabulated equations of state or compute the adiabatic index in the simulation. Instead, we adopt the standard value of 5/3, which is widely used in astrophysical contexts (and probably a good approximation on timescales after the laser irradiation). More sophisticated equations of state are often necessary to accurately describe laboratory plasmas. In our numerical setup, $\gamma$ primarily determines the density compression ratio, which, in principle, can affect the ion spectra. To assess the robustness of our results, we tested an alternative value of $\gamma = 4/3$ for the simulation labeled Run A in Table \ref{Table}, and found that it does not alter the resulting ion spectra. Although the choice of $\gamma$ remains an assumption, the shocks in our simulations are largely governed by the ion dynamics, and therefore we do not expect substantial changes when varying its value.

The sonic Mach number is defined as $M_s \equiv v_{sh} / c_s$, where $c_s = \sqrt{2 \gamma k_B T / m}$ is the adiabatic sound-speed, $k_B$ is the Boltzmann constant, and $T \equiv T_i = T_e$ is the plasma temperature, assuming ions and electron are initially in thermal equilibrium\footnote{We note that in laboratory experiments particle equilibration does not always hold. In this case, one should use the transformation $T \longrightarrow T_i + ZT_e$, where $T_i$ and $T_e$ are the ion and electron temperatures, when calculating the ion-acoustic sound speed.}. The Alfvénic Mach number is defined as $M_A \equiv v_{sh} / v_A$. The Alfv\'enic and sonic Mach numbers are related to the plasma-$\beta$ parameter by
\begin{equation}\label{eq:beta}
    M_A = \left( \frac{\gamma \beta}{2} \right)^{1/2} M_s.
\end{equation}
Since the $M_{A,s}$ usually reachable in laboratory experiments is in the range of $2-30$, we focus on this regime and test different dimensionalities and values of $\beta$, in order to track the amount of accelerated particles as a function of these two parameters and the dependence of the result on the dimensionality.

An important caveat to the numerical implementation comes from the frames of reference typically used in the laboratory and in simulations. In the laboratory, typically the upstream is at rest, whereas \texttt{dHybrid} utilizes the downstream frame of reference. In this paper, we use the laboratory/upstream frame of reference, denoted by the superscript $(u)$, to describe physical quantities in that frame of reference and/or evaluated there. The Mach numbers calculated in the downstream frame of reference, and used in the code as inputs, which are denoted by the superscript $(d)$, can be converted to the laboratory using

\begin{equation}
    M_{A,s} \equiv M_{A,s}^{(u)} = \frac{R}{R-1} M_{A,s}^{(d)},
\end{equation}
where $R\equiv n^{(d)}/n^{(u)}$ is the shock compression ratio. In contrast, $\beta$, which scales with the ratio between $M_A$ and $M_s$ through equation (\ref{eq:beta}), remains unchanged under a reference frame transformation.

We define the acceleration efficiency $\varepsilon$ as the fraction of post-shock energy density in ions with energies\cite{caprioli+14a} $\geq 10 E_{sh}$. Table \ref{Table} summarizes of the simulation parameters we used (in the laboratory/upstream frame of reference), together with their corresponding acceleration efficiency, and energy spectral index $\alpha$ at $t=10$ $\omega_c^{-1}$. We conducted a parametric study starting from $M_A=25$ and $\beta=2$ (Run A). We note that this corresponds to the lowest $M_A$ tested in Orusa \& Caprioli\cite{orusa+23}. Runs B and C are slightly less magnetized cases with $M_A = 25$, and $\beta=5$ and $18$, respectively. Finally, Run D investigates $M_A=19$ and $\beta=2$. 

All these conditions (Runs A through D) were simulated both in 2D and 3D. The field is oriented along the $y$-axis. In the three dimensional cases, the $z$-axis domain was to 20 $d_i$. We use 10 cells per $d_i$ in each direction and 8(4) ion particles per cell (ppc) in 3D(2D). To confirm that 3D particle acceleration is a genuine physical effect tied to the presence of the full third dimension, and not an artifact of particle statistics, we verified that 2D simulations with 64 and 121 ppc do not develop a non-thermal tail for a shock with $M_A = 25$ and $M_s = 19$, a regime where 3D simulations exhibit significant acceleration. We present and discuss on the values of $\varepsilon$ and $\alpha$, and their differences in 2D vs. 3D simulations, in Sec. \ref{sec:results}.

\begin{table}[t]
\begin{center}
\caption{Summary of the simulated parameters in 3D at $t= 10 \omega_c^{-1}$: Alfv\'enic Mach number, plasma-$\beta$, and sonic Mach number; together with inferred parameters of interest: acceleration efficiency $\varepsilon$, compression ratio $R$, and energy spectral index $\alpha$ . In all simulations, the initial magnetic field inclination was fixed to $\thbn=90\deg$. No accelerated particles are found in the corresponding 2D simulations.}
\label{Table}
\begin{tabular}{ l c c c | c c c}
 \hline \hline
Run &\quad $M_A$\quad &\quad $\beta$\quad &\quad $M_s$ &\quad $\varepsilon$ ($> 10E_{sh}$) &\quad $R$ &\quad $\alpha$ \\
\hline 
A   &   $25$    &   $2$  & $19$  & $0.3\% $ & $4.2$ & $ 5.4$  \\
B   &   $25 $   &   $5$  & $13$ & $0.2\%$ &  $4.3$ & $5.7$\\
C   &   $28$    &   $18$ & $7$  & $0.05\%$  & $3.5$ & $8$\\
D   &   $19$    &   $2$  & $15$ & $0.04\% $& $4.3$ & $8$\\
 \hline \hline
\end{tabular}
\end{center}
\end{table}

\begin{figure*}
    \centering
    \includegraphics[width=16.5cm]{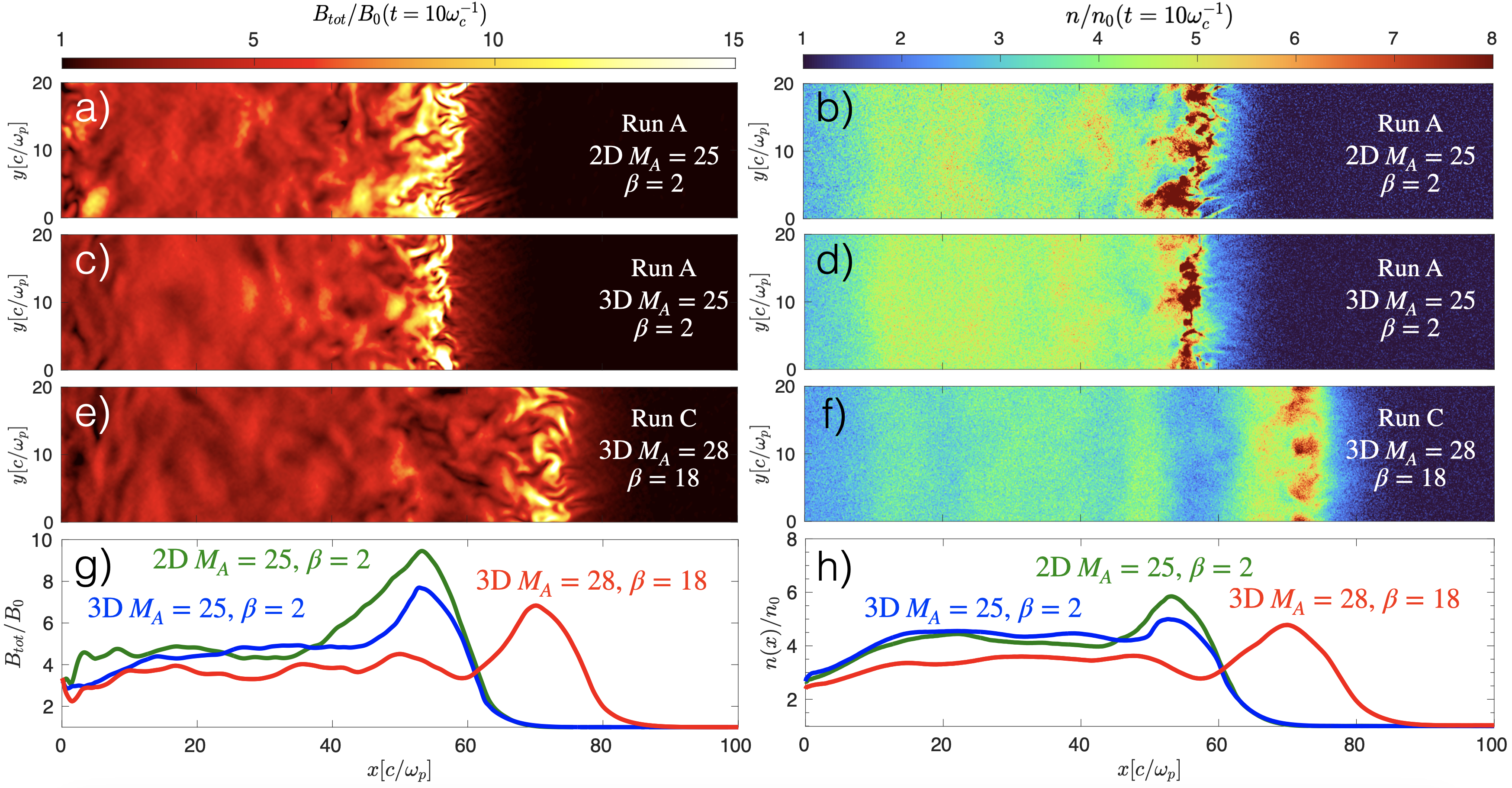}
    \caption{Simulated magnetic field and density (normalized by upstream parameters) at $t= 10\omega^{-1}_c$ for indicated conditions $(M_A, \beta)$, in 2D and 3D. In the latter case, the colormap corresponds to a slice through the mid-plane of the simulation. Panels (a), (c), (e): magnetic field. Panels (b), (d), (f): density. Panels (g) and (h) are integrated magnetic fields and density for each run, respectively.}
    \label{fig:shock structure}
\end{figure*}

\section{Numerical results}
\label{sec:results}
We present the simulation results, splitting it in different aspects of the physics of interest. First, we discuss the differences in shock structure for different values of $(M_A,\beta)$. Second, we present the calculated ion spectra and the relation between the relevant parameters, dimensionality, and the emergence (or not) of a non-thermal tail. Third, we will show the evolution of the most energetic ions found in the simulations, which further highlights the importance of dimensionality to ion acceleration.


\subsection{Shock structure}
The general structure of a quasi-perpendicular collisionless shock is well known\cite{marcowith+16}. Quasi-perpendicular shocks exhibit a density gradient, the shock front, called the \emph{ramp}. Ions accumulate behind the ramp, generating an \emph{overshoot} in the magnetic field. Moreover, the shock front reflects incoming ions back into the upstream, forming a slightly denser region ahead of the ramp known as the \emph{foot}.  This general behavior is observed in both 2D and 3D. However, the strength of the overshoot, together with length-scales related to the ramp and the foot, can depend on $M_A$, $\beta$, and the dimensionality of the system.

Beyond the one-dimensional description of the shock, these systems exhibit strongly fluctuating density and magnetic components. The density and magnetic structures for different Runs and dimensionality at $t=10 \omega_c^{-1}$ are shown in Figure \ref{fig:shock structure}. Panels (a)$-$(d) show the case that most efficiently accelerates ions $(M_A=25,\beta=2)$ in 2D and 3D. Filamentary structures are visible in the ramp and foot. The plasma conditions are in the intersection between Alfv\'en ion cyclotron- and the ion-Weibel-dominated unstable regime, hence the emergence of filaments can be attributed to either of these instabilities\cite{nishigai+21,matsumoto+15,bohdan+21,jikei+24}.As incoming ions encounter the population of ions reflected by the shock, counter-streaming beams are established in the upstream region. This configuration is unstable to the ion Weibel instability, which leads to the formation of small-scale current filaments oriented along the shock normal. These filaments induce transverse magnetic fields, perpendicular to both the shock normal and the upstream magnetic field. As the shock advances, it compresses these structures, amplifies and advects them into the downstream region. There, the filaments merge and evolve into a turbulent magnetic field, which dominates the downstream region. In both 2D and 3D simulations, the density is compressed by the shock, with an overshoot immediately behind it that eventually relaxes into a weakly turbulent state dictated by the standard compression ratio of 4. We emphasize that, despite the fact that two cases look very similar visually, the out-of-plane structure of the shock is the key for ion acceleration\cite{orusa+25}. The results from case $(M_A=25, \beta=5)$ are similar to panels (a)$-$(d) and are not presented for conciseness.

Figure \ref{fig:shock structure}e and f show the shock structure at $(M_A=28, \beta = 18)$ and is therefore less hypersonic with $M_s = 7$ than the Runs discussed above. This case provides less insight into ion acceleration and it is relevant for shocks in the heliosphere. The dominance of thermal pressure over magnetic pressure suppresses the development of turbulence at kinetic scales relevant for ion injection in the downstream region, resulting in a more laminar flow. In fact, when the upstream plasma beta is $\beta \gg 1$, the influence of the magnetic field on the shock jump conditions becomes negligible \cite{tidman+71}. The density and magnetic compression ratio is closely tied to the sonic Mach number, with an observed $R=3.5$, instead of 4 (the expected value for strong shocks) in the far downstream region. This value of $R$ is consistent with predictions based on the Rankine-Hugoniot conditions for a weakly magnetized shock\cite{guo+14a}, which explains the displacement of the shock position relative to other cases: lower compression implies that the shock forms and propagates more rapidly. Similarly, the overshoot immediately behind the shock is weaker than the Runs with larger values of $M_s$. Notice that this simulation was performed with the same value of $M_A^{(d)}$ in the downstream reference frame as the other simulations, but due to the lower compression ratio, this results in a higher value of $M_A$ in the laboratory frame. The final case with $(M_A=19,\beta=2)$ exhibits lower amplitude magnetic fluctuations and amplification with respect to $(M_A=25,\beta=2)$, since they approximately scale with $\sim \sqrt{M_A}$ (see Refs. \onlinecite{kato+10, bohdan+21, matsumoto+15}.)

An important piece of analysis is averaging the simulations in the $yz$-plane to study the characteristic 1D structure of the shock in each case. The results are presented in Figure 1g and h. They show that in the same conditions, 2D simulations exhibit a slightly higher overshoot compared to 3D of order $10\%$ with a sharper transition into the downstream in the latter case. Nevertheless, the downstream density and magnetic field are equal. The simulation at higher $\beta$ propagates faster and exhibits a lower amplitude. As discussed above, the compression ratio is also lower than the more magnetized cases.

\subsection{Ion energy spectra}

\begin{figure}
    \centering
    \includegraphics[width=8cm]{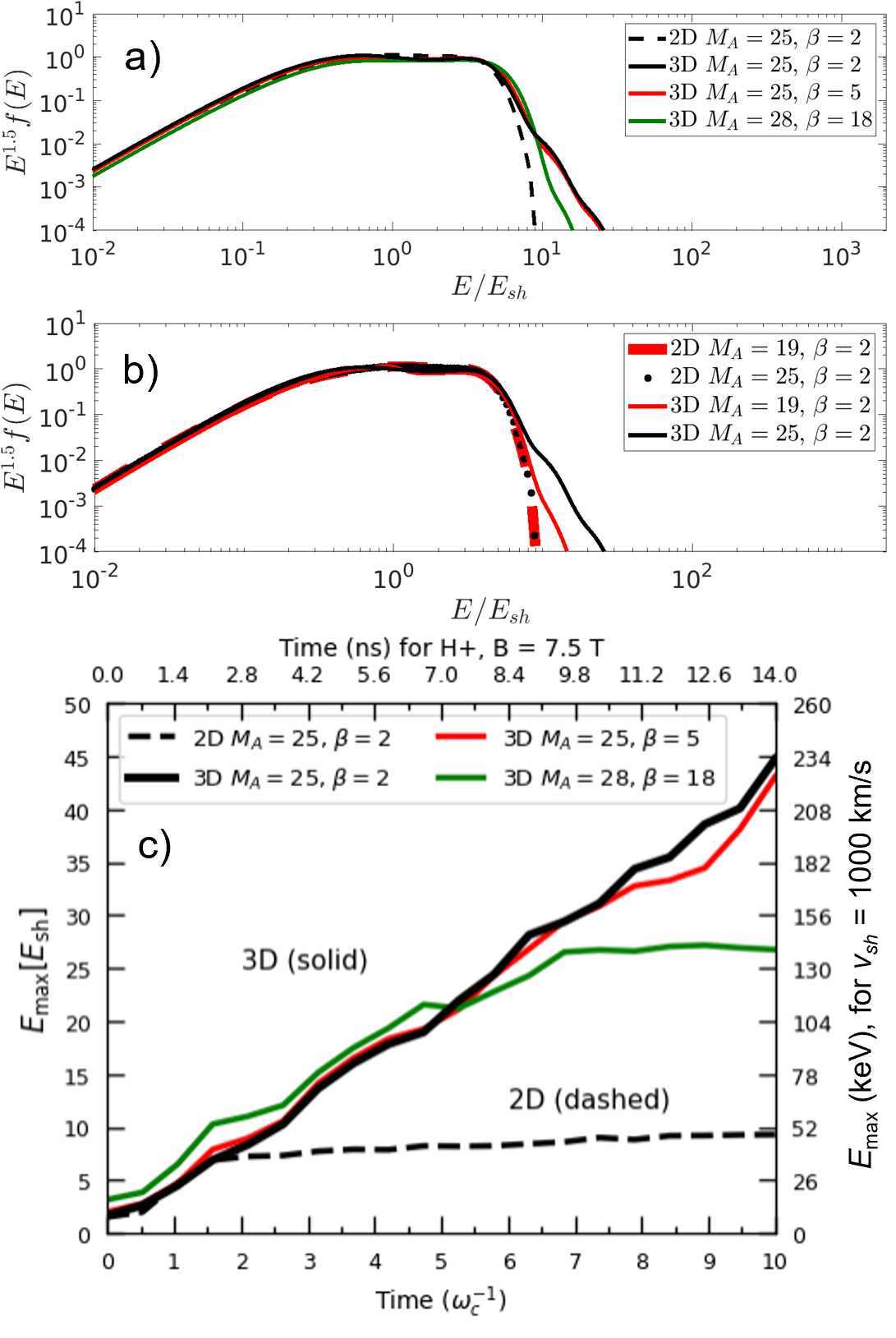}
    \caption{Characterization of accelerated ions. a) Energy spectra for different conditions at $t=10\omega_c^{-1}$. b) Comparison of ion spectra between 2D and 3D in two specific conditions.  (c) Time evolution of the maximum energy of ions at different conditions.}\label{fig:energy spectra}
\end{figure}

As mentioned before, despite the visual similarity of structures between 2D and a slice of a 3D simulations, there are notable differences in the spectrum or accelerated ions. Figure \ref{fig:energy spectra}a shows the energy spectra of ions for different regimes of $(M_A,\beta)$ and dimensionality, as a function of particle energy normalized by the energy per ion moving at shock speed. Notice the convergence of the thermal and supra-thermal population with $E \lesssim 5 E_{sh}$, consisting of particles that are either advected downstream or reflected once, completing at most a single gyration upstream before being carried into the downstream region. However, for $E\gtrsim 10E_{sh}$ there are appreciable differences. First, in 3-dimensions, the spectral tail above $10 E_{sh}$ for the cases $(M_A =25, \beta=2)$ and $(M_A=25,\beta=5)$ is remarkably similar with a spectral index $\alpha \approx 5.5$ (see Table \ref{Table} for precise values). For these two cases, the magnetic field structure is very similar, and the probability of advection into the downstream region is nearly the same, resulting in an almost identical spectrum. 

The collisionless shock in the case $(M_A=28,\beta=18)$ also develops a softer non-thermal tail compared to the more hypersonic case, with spectral index $\alpha = 8$. In this case, the dominance of thermal pressure over magnetic pressure inhibits the development of turbulence at kinetic scales relevant for ion injection in the downstream region, thereby increasing the likelihood of particle advection.

The simulated spectra in 2D does not exhibit the development of a non-thermal tail (in any condition), hence the ion acceleration is enabled only by the dimensionality of the system. This is further shown in Figure \ref{fig:energy spectra}b, which presents the ion spectra in 2D and 3D for two different conditions. As opposed to the case $M_A = 25$, when $M_A = 19$ the non-thermal tail is less pronounced. This is because this Alfvénic Mach number falls within the threshold region for ion injection. Since the level of downstream magnetic field amplification scales approximately as $\sqrt{M_A}$, the reduced turbulence increases the likelihood of particle advection.

The spectra shown in Fig. \ref{fig:energy spectra}a and b also lead to different maximum ion energies, $E_{\rm max}$. We can further see differences between 2D and 3D simulations by investigating the evolution of maximum particle energy in the simulation, which is shown in Figure \ref{fig:energy spectra}c. We found that the maximum energy increases linearly only in 3-dimensions. Moreover, the cases $(M_A =25, \beta=2)$ and $(M_A=25,\beta=5)$ exhibit very similar maximum ion energies over time, reaching $E_{\rm max} \sim 40 E_{sh}$ at 10 $\omega_c^{-1}$. The right hand side axis shows the hypothetical equivalent maximum particle energy to be observed in the laboratory for a shock propagating at 1000 km/s, which would accelerate particles to energies on the order of 200 keV.

In contrast, for the 2D ($M_A=25, \beta=2$) case, particles barely exceed 10 $E_{sh}$, saturating in a few gyro-periods. Finally, for $M_A=28$ and $\beta=2$, ions undergo at most a few gyrations, leading to the saturation of the maximum energy over time, as seen in Fig. \ref{fig:energy spectra}c, with a final $E_{sh} \approx 25$.



\subsection{Acceleration efficiency}
\begin{figure}
    \centering
    \includegraphics[width=8.75cm]{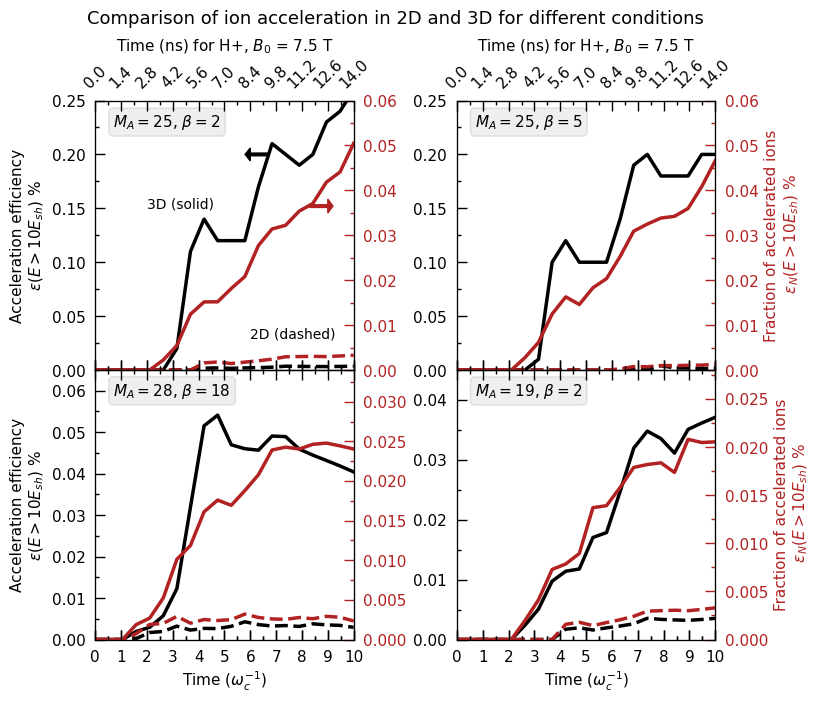}
    \caption{Evolution of the acceleration efficiency and the percentage fraction of ions with final energy $\geq 10 E_{sh}$ in 2D and 3D for indicated conditions $(M_A,\beta)$.}
    \label{fig:efficiency}
\end{figure}

The differences in spectra between 2D and 3D, as well as variations in $M_A$ and $\beta$, directly translate into differences in the acceleration efficiency $\varepsilon$ and the percentage fraction of accelerated ions with final energy $\geq 10 E_{sh}$. These quantities are presented in Figure \ref{fig:efficiency} for the different runs performed in both 2D and 3D. The efficiency $\varepsilon$ is represented by a black line, while the percentage fraction of accelerated ions is shown with a red line. Dashed lines correspond to the 2D setup, whereas solid lines represent the 3D case.

Similar to the maximum particle energy study, we consistently find that 3D simulations allow a higher fractions of particle to be accelerated, i.e. ion acceleration is suppressed in 2-dimensions. However, the acceleration efficiency drastically varies depending on the $(M_A,\beta)$ of the system. For the case ($M_A=25, \beta=2$), reported in Fig. \ref{fig:efficiency}a, we obtain $\varepsilon \sim 0.3\%$ in 3D, which does not reach saturation within $\sim 10 \omega_c^{-1}$. This results in a percentage fraction of accelerated ions at the 0.05\% level. In contrast, the 2D simulation yields $\varepsilon \sim 0.01$. Coherently with the measured spectrum, similarly, for ($M_A=25, \beta=5$), a difference between 2D and 3D simulations is observed, as reported in Fig. \ref{fig:efficiency}b, with $\varepsilon \sim 0.2\%$ and fraction of accelerated ions of 0.05\%, comparable to the value obtained for $\beta=2$.

For $(M_A=28,\beta=18)$, ions undergo at most a few gyrations, leading to the saturation of $\varepsilon$ over time, as shown in Fig. \ref{fig:efficiency}c. In this case, the differences between 2D and 3D are smaller, with $\varepsilon \sim 0.05\%$, approximately a factor of 6 lower than in the ($M_A = 25$, $\beta = 2$) case—although some acceleration is still observed. A similar $\varepsilon$ and fraction of accelerated ions are obtained for ($M_A=19, \beta=2$), as reported in Fig. \ref{fig:efficiency}d, decreasing values compared to the higher $M_A$ and lower $\beta$ cases. These two cases serve as examples of weak, but nonzero, acceleration.


\section{Discussion}

\subsection{Parameter space for ion acceleration in laboratory quasi-perpendicular shocks}

From the simulation campaign presented here, we identify three distinct regions in the parameter space of ($M_A, M_s)$ relevant to laboratory experiments, defined by their particle acceleration efficiency.

\begin{itemize}
    \item \emph{Strong acceleration}: For $M_A\gtrsim25$ and $M_s\gtrsim13$, conditions are highly favorable for particle acceleration, as indicated by the presence of a non-thermal tail in the energy spectrum with efficiencies reaching approximately 0.2\%. The maximum particle energy at $t=10\omega_c^{-1}$ is $E_{\text{max}}\approx 40 E_{sh}$. In this regime, fully capturing 3D effects is essential for an accurate description of the accelerated ion spectra.
    \item \emph{Weak acceleration}: For $19\lesssim M_A< 25$ and $M_s\gtrsim7$. There are differences between 2D and 3D simulations, but not as pronounced as in the strong acceleration case, meaning that the high-energy tail is probably challenging to detect experimentally. In fact, only a small fraction of particles undergo non-thermal acceleration, with energy efficiencies around 0.05\%, reaching a maximum energy of $E_{\text{max}}\approx 25 E_{sh}$. While this regime is less efficient for particle acceleration, it is more accessible for laboratory experiments.  
    \item \emph{No acceleration}: For $M_A < 19$ and $M_s < 7$, the particle spectra from 2D and 3D simulations are indistinguishable, and particles can gain at most $E_{\rm max} \sim 10 E_{sh}$. In this regime, particles are typically reflected only once by the shock before being advected away, rather than crossing the shock front multiple times, preventing sustained acceleration. 
\end{itemize}

Additionally, we identify two distinct time intervals in the evolution of the shock: during the first 5 $\omega_c^{-1}$, the shock forms, and a small population of energetic particles emerges, with acceleration efficiencies already exceeding zero. From 5 to $10\omega_c^{-1}$, the shock continues to develop, leading to a progressive increase in both the energy efficiency and the maximum energy of the accelerated particles. These temporal conditions further constrain the emergence of non-thermal ions in laboratory experiments, since the shock must be sufficiently long-lived such that these processes can occur. We quantify all of these requirements in the next section.

\subsection{Scaling criteria}\label{sec:scaling}

\begin{figure*}[t]
    \includegraphics[width=0.49\textwidth]{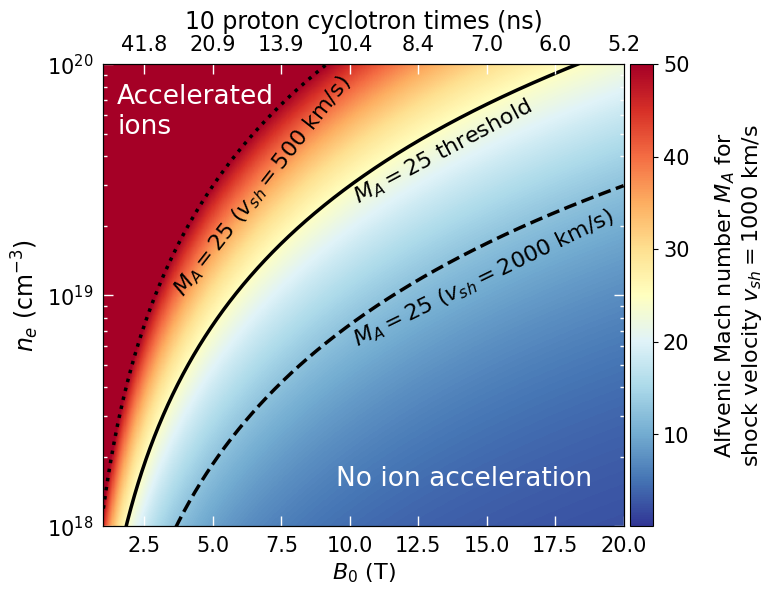}
    \includegraphics[width=0.49\textwidth]{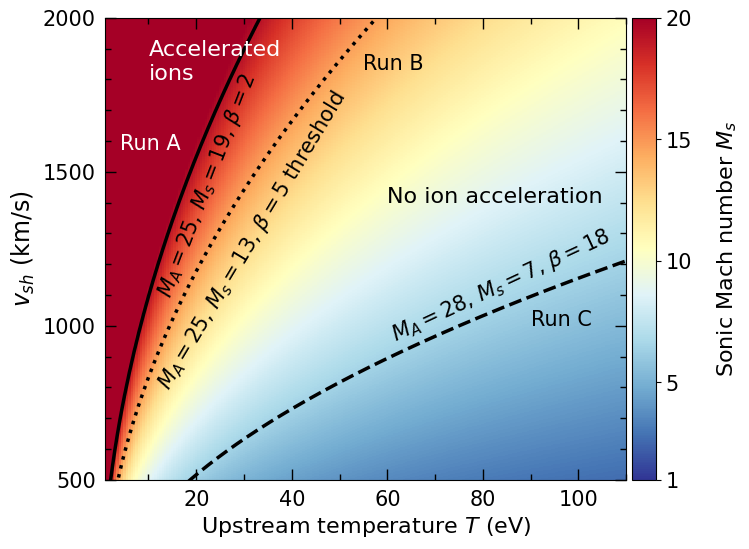}
    \caption{Left panel: Phase diagram of $M_A$ as a function of the $(B_0,n_e)$-space, for $v_{sh}=1000$ km/s. The solid black line refer to the Alfv\'enic locus $M_A=25$, which defines the threshold for strong proton acceleration obtained in Sec. \ref{sec:results}, such that $v_{sh} = 1000$ km/s. The diagram also shows different Alfv\'enic loci (dashed and dotted lines) defining the ion acceleration threshold covered by different shock velocities. At higher shock velocities, the accessible parameter space expands, allowing larger $M_A$ values and enhanced particle acceleration. Right panel: Phase diagram of $M_s$ in $T-v_{sh}$-space. The black lines (solid, dotted, and dashed show the sonic loci covered by by Runs A, B and C, respectively. The dotted black line correspond to the \emph{sonic locus} above which we can have large ion acceleration.}
    \label{fig:parameters_space}
\end{figure*}

Our simulations elucidate conditions that can be achieved in current laboratory experiments, establishing the threshold for ion acceleration. These results provide a foundation for discussing how the parameter space evolves under different scaling conditions. 
This section has the goal to quantify what parameters (such as particle species, external magnetic field, upstream density, upstream temperature, shock velocity) are needed in the laboratory to produce 3D ion acceleration. Using scaling considerations we will show that it is plausible to control the acceleration process in the laboratory.

We first explore the parameter space $(M_A, M_s)$ in realistic laboratory conditions. The left panel of Figure \ref{fig:parameters_space} shows the phase diagram of $M_A$ as a function of an externally applied $B_0$ and upstream electron density $n_e$ (the ion density is straight-forward to calculate from quasi-neutrality $n=n_e/Z$), assuming a shock velocity of $v_{sh} = 1000$ km/s in an electron-proton plasma. The solid line represents the \emph{Alfv\'enic locus}, i.e. the combination of $B_0$ and $n_e$ required to achieve $M_A = 25$ (at a given $v_{sh}$) which, based on our previous results, defines a threshold for efficient particle acceleration. In addition, we include two alternative cases for $M_A=25$: a dotted line for $v_{sh} = 500$ km/s and a dashed line for $v_{sh} = 2000$ km/s. The upper horizontal axis indicates the equivalent of $10 \omega_c^{-1}$ in nanoseconds for a proton (the lightest of ions), providing insight into the temporal constraints of different values of $B_0$. The results show that the shock should be sustained for $\gtrsim 10$ ns to allow ion acceleration. Below we will find scaling criteria for any other ion species considered. Naturally, increasing $B_0$ requires a corresponding increase in $n_e$ to maintain the required $M_A$, but it also increases the number of captured $\omega_c^{-1}$, which plays a crucial role in the acceleration process. Nevertheless, for $v_{sh} = 1000$ km/s, a significant region of the parameter space satisfies the conditions necessary for ion acceleration and the higher is $M_A$ the larger is the acceleration. For greater shock velocities, the available area above the locus increases and even higher $M_A$ could be obtained. The current yellow region in Fig. \ref{fig:parameters_space}(left) would shift to where the dashed line is for $v_{sh}=2000$ km/s, and for the values of $B_0$ and $n_e$ shown, that would allow access to even higher $M_A$.

It is important to quantify the change of the threshold Alfv\'enic locus when different shock speeds are considered. Two calculations for $v_{sh} = 500$ km/s and $v_{sh} = 2000$ km/s are presented in dotted and dash lines, respectively. Reducing the shock velocity to $v_{sh} = 500$ km/s significantly limits the $(B_0,n_e)$-space where acceleration can occur. Conversely, increasing the velocity to $v_{sh} = 2000$ km/s expands the viable parameter range, making it easier to sustain a high-$M_A$ shock for an extended duration, which is beneficial for efficient particle acceleration. We provide scaling considerations for the shock velocity below.

In the right panel of Fig. \ref{fig:parameters_space}, we show the phase diagram for $M_s$ as a function of upstream temperature $T$ and $v_{sh}$. The dotted black line correspond to the \emph{sonic locus} covered by our simulations discussed in the previous section, above which we can have large ion acceleration. When the upstream material is initially cold ($T\approx 10$ eV), the shock velocity required to accelerate ions is greatly relaxed in the $\beta=2$ case, compared to systems at higher plasma-$\beta$. 


We now introduce scaling conditions that constrain the emergence of efficient ion acceleration in three dimensions. From these conditions we can identify experimental configurations in which the upstream plasma material (characterized by atomic weight $A$ and charge state $Z$), the upstream magnetic field $B_0$, density $n_e$, and the upstream temperature $T$ can be selected to enable (or suppress) 3D ion acceleration. Based on our numerical simulations, the following criteria must be met:

\begin{itemize}
    \item[1.] The shock must be highly super-Alfvénic, with $M_A \gtrsim M_{A,\text{crit}} = 25$.
    \item[2.] The shock must have a low to moderate plasma-$\beta \lesssim 5$. Given that the shock is already highly super-Alfvénic, this condition translates to a sonic Mach number of approximately the same order as $M_A$, specifically $M_s \gtrsim M_{s,\text{crit}} = 13$.  
    \item[3.] Ions must be accelerated to energies exceeding $10 E_{sh}$ after the shock has been driven for at least $N \gtrsim  t/\omega_c^{-1} \gtrsim 5$. However, to ensure a significant number of accelerated particles, it is preferable to sustain the shock for at least $N \gtrsim N_{\text{crit}} = 10$.   
\end{itemize}

To achieve these requirements\footnote{An obvious additional requirement is that the ion-ion mean free path is much larger than the density gradient length-scale. It has been demonstrated that laboratory experiments can achieve this, so we will assume it to be the case, but one should check when confronting an actual experimental configuration.}, we begin by noting that the first step—the choice of how many ion cyclotron times we aim to achieve—depends only on the magnetic field. For the shock to develop for $N_{\text{crit}} = 10$ within the experimental time-frame $\tau_{\text{exp}}$, the upstream magnetic field must satisfy:  
\begin{equation}\label{eq:min_B}
    B_0 = \left(\frac{A}{Z}\right) \left(\frac{m_p}{e} \right) \frac{N_{\rm crit}}{\tau_{\rm exp}} \approx 10.4 \left(\frac{A}{Z}\right) \left(\frac{\rm 10 \; ns}{\tau_{\rm exp}} \right) \; {\rm T} ,
\end{equation}
where, in the last approximation, we have used the ratio of the proton mass $m_p$ to the fundamental charge $e$ to derive a practical expression for the magnetic field in Teslas, with $\tau_{exp}$ expressed in nanoseconds. Setting $\tau_{exp} = 10$ ns results in a required field of 10.4 T to achieve $N_{\text{crit}} = 10$.

Second, the maximum upstream temperature requirement is linked to the shock velocity through the condition on the sonic Mach number $M_s$. For a given shock velocity and $M_s \geq M_{s,\text{ crit}}$, the upstream temperature $T^*$ must satisfy:  
\begin{align}
    T^*&= \left( \frac{A}{1+Z} \right) \left( \frac{m_p}{\gamma k_B}\right)  \left(\frac{v_{sh}}{M_{s,\text{ crit}}}\right)^2 \\ \nonumber
    &\approx 50 \left( \frac{2A}{1+Z} \right)   \left(\frac{v_{sh}}{1650  \; \rm km/s}\right)^2 \rm eV \\ 
\end{align}

where we have assumed $\gamma = 5/3$, $M_{s,\text{ crit}}=13$, and expressed $v_{sh}$ in km/s and the temperature in electronvolts to derive the practical formula above.  

Third and finally, the upstream magnetic field $B_0$ and density $n_e$ determine the Alfvén velocity $v_A$ in the upstream region. Under these conditions, the shock speed $v_{sh}^\ast$ sets the Alfvénic Mach number. The requirement $M_A \geq M_{A,\text{ crit}} = 25$ implies that the upstream density must exceed a certain threshold, given by the condition below, assuming $N_{\rm crit} = 10$:  
\begin{align}\label{eq:max_ne}
    n_e^\ast &= \left(\frac{A}{Z} \right) \left( \frac{m_p}{\mu_0e^2} \right) \left( \frac{N_{\rm crit}}{\tau_{\rm exp}} \right)^2 \left(\frac{M_{A,\text{ crit}}}{v_{sh}}\right)^2 \\ \nonumber
    & \approx 35\times 10^{18}\left(\frac{A}{Z} \right) \left( \frac{\rm 10 \; ns}{\tau_{\rm exp}} \right)^2 \left( \frac{\rm 1000 \; km/s}{v_{sh}} \right)^2 \; {\rm cm}^{-3}.
\end{align}
where $\tau_{\rm exp}$ is in ns, $v_{sh}$ is in km/s, and $n_e^\ast$ cubic centimeters.  

\begin{figure}
\centering
    \includegraphics[width=0.49\textwidth]{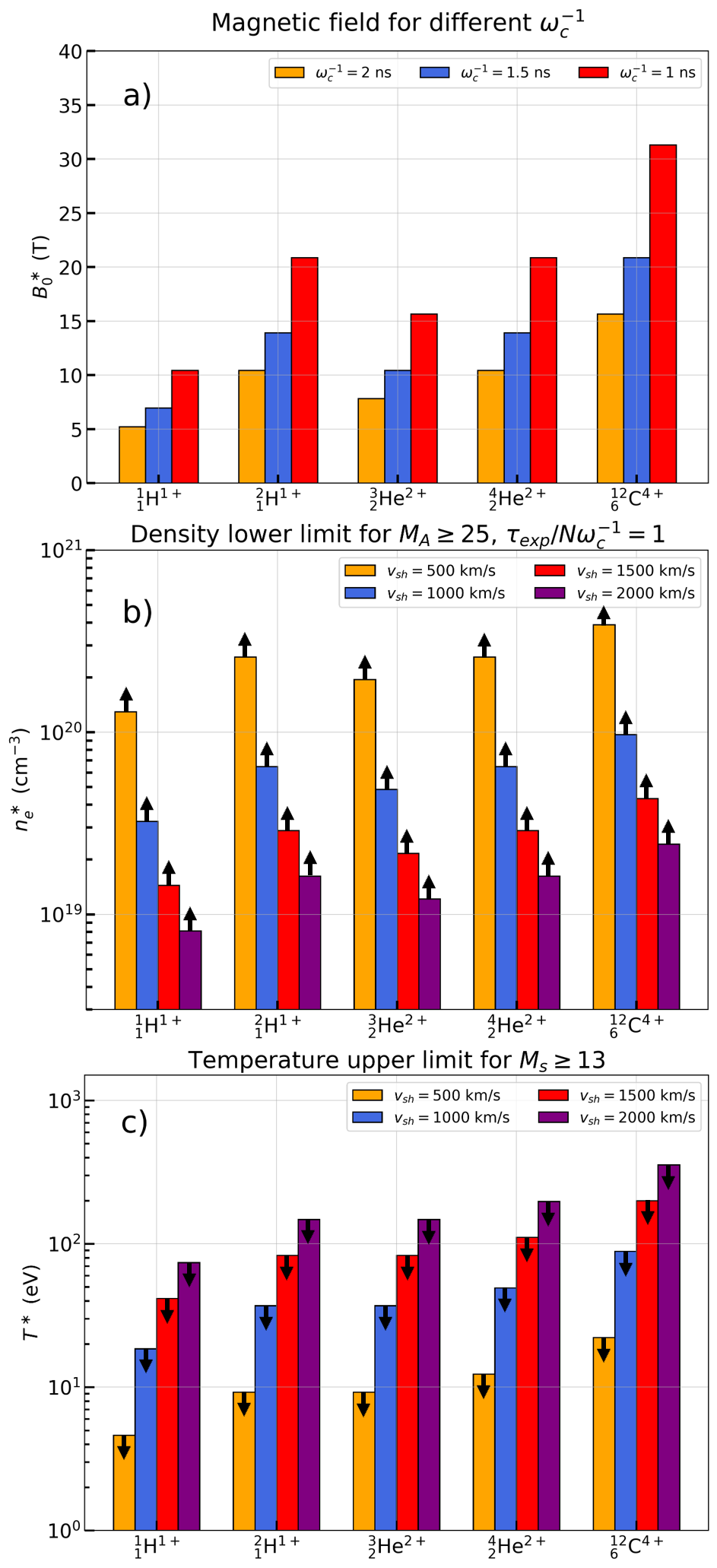}
    \caption{Dependence of the required $B_0$, $n_e$, and $T$ to achieve $M_A \gtrsim 25$ and $M_s \gtrsim 13$ for different elements, isotopes, and charge states. a) Calculated upstream magnetic field $B_0$ such that the ion cyclotron time is 1 ns (red diamond), 1.5 ns (black square), and 2 ns (green circles). b) Lower limit of $n_e$ for different shock velocities (green circles: 500 km/s, magenta squares: 1000 km/s, black diamonds: 1500 km/s, and red hexagons: 2000 km/s), assuming $\tau_{exp}/N_{\text{crit}} \omega_c^{-1} = 1$. Upward pointing arrows emphasize that, for ion acceleration to occur, the upstream density must be larger (or equal) than this value. c) Upper limit of the temperature required to reach $M_s \gtrsim 13$ for various $v_{sh}$. Downward pointing arrows emphasize that, for ion acceleration to occur, the upstream temperature must be lower (or equal) than this value.} 
    \label{fig:scaling}
\end{figure}

Equations (\ref{eq:min_B}) through (\ref{eq:max_ne}) can be used to design experiments where 3D effects are either significant or negligible for ion acceleration. In practice, the magnetic field can be externally imposed using inductive coils driven by a specific voltage, while the shock velocity and experimental time frame can be controlled by selecting an appropriate laser driver (intensity, duration, and total energy). The density can be adjusted using a pressurized gas jet or a cross-wind plasma. On the other hand, the upstream temperature is much harder to control, in particular to cool down (one can use an auxiliary heater beam to raise the upstream temperature, for example). 

In Figure \ref{fig:scaling}, we illustrate how the required values of $B_0$, $n_e$, and $T$ for ion acceleration depend on the plasma composition, based on the equations presented above. The ion species were selected because they are generally light and available in gas form. The heaviest ion species considered is carbon, which is present when shooting plastic targets, so it might be of interest to experimentalist. Moreover, we have assumed that carbon ions have a charge state of $Z=4$, consistent with ionization tables\cite{Chung2005} in the range $10$ eV $\leq T \leq 90$ eV and electron density $n_e = 10^{18}$ cm$^{-3}$, which are typical conditions in laboratory experiments. Figure \ref{fig:scaling}a shows the magnetic field required to obtain $\omega_c^{-1}$ of 1 ns, 1.5 ns, and 2 ns for different ionic species. This allows calculating a value of $B_0$ such that the upstream ions gyrate $N$ times in a given experimental time frame $\tau_{\rm exp}$. Notice that in all cases ions can gyrate on single-nanosecond scales with fields $< 40$ T. Indeed, for experiments with light ions (such as hydrogen and helium), this magnetic field is $< 15$ T, which can be applied using current pulsed-power capabilities, such as the Magneto-Inertial Fusion Electrical Discharge System (MIFEDS\cite{Barnak2018}) on the OMEGA laser.

Assuming $\tau_{exp}/N_{\rm crit} \omega_c^{-1} =1$ (i.e. the experiment always achieves the critical number of ion gyrations), we can determine the corresponding lower limit for the electron density required to achieve the desired $M_A$ for different shock velocities $v_{sh}$. Figure \ref{fig:scaling}b shows values of $n_e^\ast$ for different ion species and shock velocities. For $v_{sh} < 500$ km/s, we find that typically $n_e^\ast > 10^{20}$ cm$^{-3}$, regardless of the ion species. As a point of reference, the gas jet nozzles at the Laboratory for Laser Energetics\cite{Mcmillen2024} can achieve gas densities of few $\times 10^{19}$ cm$^{-3}$, making it challenging to have a dense enough upstream with such a low velocity (not to mention that the system could become collisional). Cross-wind plasmas driven by a secondary beam are one order of magnitude more dilute\cite{schaeffer+19}. The requirements are more easily met for higher shock speeds $v_{sh}>1000$ km/s, in particular for proton-electron plasmas.

Finally, as we mentioned above, the upstream temperature constraints the minimum shock velocity such that the system is hypersonic enough ($M_s \gtrsim 13$) to accelerate ions. Figure \ref{fig:scaling}c shows the maximum upstream temperature for a number of ion species and shock velocities. For most materials and speeds, $T^\ast < 100$ eV, which seems reasonable for an unperturbed upstream plasma.




Our results show that, for a given configuration of magnetic field, density, and shock velocity, the upstream plasma composition can be selected as a switch to enable or suppress ion acceleration. This is particularly useful experimentally, as it can be achieved simply by replacing the gas cylinder in a pressurized gas jet or changing the target material. We show this more explicitly below. In the regime relevant to laser-driven experiments, these requirements can be summarized as the scaling hierarchy
\begin{equation}\label{ineq:dimless}
   N_{\text{crit}} \lesssim \frac{\tau_{exp}}{\omega_c^{-1}} \lesssim M_s \lesssim M_A.
\end{equation}

Notice that, in practical terms, these conditions lead to an optimization problem. For example, an experimenter may try to increase the magnetic field to decrease the ion gyro-period. However, all other things being equal, this would also decrease $M_A$. It is then useful to calculate if a particular configuration such that a criterion for ion acceleration can be satisfied. The strategy is to establish scaling requirements for ion acceleration in three dimensions, beginning with an electron-proton plasma under the assumption that it is fully ionized (i.e., $Z_p = 1$ and $A_p = 1$), where the subscript $p$ denotes protons, which can then be scaled to other materials (represented by different atomic weights and charge states) for which the threshold for ion acceleration can be satisfied (or not). For a given upstream magnetic field, the number of ion cyclotron periods $N$ can be expressed as the product of the ion cyclotron frequency and the characteristic experimental duration over which the shock evolves, $N \sim \omega_{c} \tau_{exp}$. Thus, the requirement for ion gyrations can be scaled from a proton plasma to heavier and/or more strongly charged ion species with atomic weight $A$ and charge state $Z$.

\begin{equation}\label{eq:ion_gyr_scale}
    \omega_{c,p}\tau_{exp} = N_p \Longrightarrow N = \frac{Z}{A}N_p
\end{equation}
Similarly, the Alfv\'enic and sonic Mach numbers scale with ion properties, respectively, as:
\begin{equation}\label{eq:sonic_Alfvenic_scale}
     M_{A} = \left(\frac{A}{Z} \right)^{1/2}M_{A,p}, \; \; \; \; M_{s} = \left(\frac{2A}{Z+1} \right)^{1/2}M_{s,p}.
\end{equation}

The hierarchy required for ion acceleration, inequalities (\ref{ineq:dimless}), along with the scaling relations (\ref{eq:ion_gyr_scale}) to (\ref{eq:sonic_Alfvenic_scale}), can be used to identify experimental configurations where ions are accelerated through 3D effects or, alternatively, to verify when a 2D simulation provides an accurate representation of the experiment. Notice that, in a given experimental configuration defined by $(B, n_e, v_{sh})$ that satisfies the inequalities (\ref{ineq:dimless}) for a given material, it is possible to find a different one that does not because of the different scaling with $(A,Z)$ of equations (\ref{eq:ion_gyr_scale}) and (\ref{eq:sonic_Alfvenic_scale}). As an example, let us consider an electron-proton ($A_1=1, Z_1=1$) collisionless shock such that it is in the strong acceleration regime (indicated by the subscript 1), with $(M_{A,1} = 25, M_{s,1} = 13)$ and is sufficiently long-lived with $\omega_{c,1}\tau_{exp} = N_1 =N_{\text{crit}}$. Using the same experimental setup (laser driver, magnetic fields, and so on), one could change the upstream material (denoted by the subscript 2) and use a different isotope of hydrogen, such as deuterium ($A_2=2, Z_2=1$). Then the system would be described by $(M_{A,2}=35,M_{s,2} = 18)$ however, $N_2 = N_{\text{crit}}/2$. Therefore the system does not have enough time to accelerate ions, which would effectively shut down the signal. In principle, for a given laser experiment, one could find interesting combinations of ion species to explore the $(M_A, M_s)$ parameter space and find different ion spectra. These results could then be compared with simulations as a means to validate numerical codes.


\subsection{Connection to current and potential future experiments}\label{sec:discussion}

As mentioned earlier, evidence of ion energization in collisionless perpendicular shocks generated in laser plasma experiments has been reported\cite{schaeffer+19,yao+21, yamazaki+22}. In this Section, we survey these experimental results with the conditions we found are relevant to 3D ion acceleration. Based on our results, we found that these experiments should be well-described by 1D and 2D kinetic simulations. We will close this discussion by proposing a few plausible parameter configurations that could be explored in future studies to further investigate ion acceleration in collisionless shocks. 

\subsubsection{Schaeffer et al. at the OMEGA laser facility}

The experiment conducted by Schaeffer et al. \cite{schaeffer+19} at the OMEGA laser facility \cite{boehly+95} marked the first laboratory observation of time-resolved electron and ion velocity distributions in magnetized collisionless shock precursors (i.e. not fully formed). A single inductive coil made of copper wires was driven using MIFEDS, producing a upstream ambient magnetic field of 10 T that was applied to pre-magnetize a single laser beam-driven cross-wind upstream plasma, filling a large volume in front of a plastic (CH) foil target. By focusing two drive beams on this target, a hypersonic piston was produced, generating a shock. Coupling these experiments with dedicated simulations\cite{Germaschewski2016,Schaeffer2017}, the authors showed that the hydrogen from the foil couples efficiently with the upstream, creating a proton-electron dominated collisionless shock.

The shock precursor propagated at a speed of approximately 750 km/s, and the authors inferred $(M_A=15, M_s=15)$ and the experimental time frame was $\tau_{exp}\approx 4$ ns, enough to sustain $\sim 4$ proton gyrations.

\begin{figure}[t]
    \includegraphics[width=0.45\textwidth]{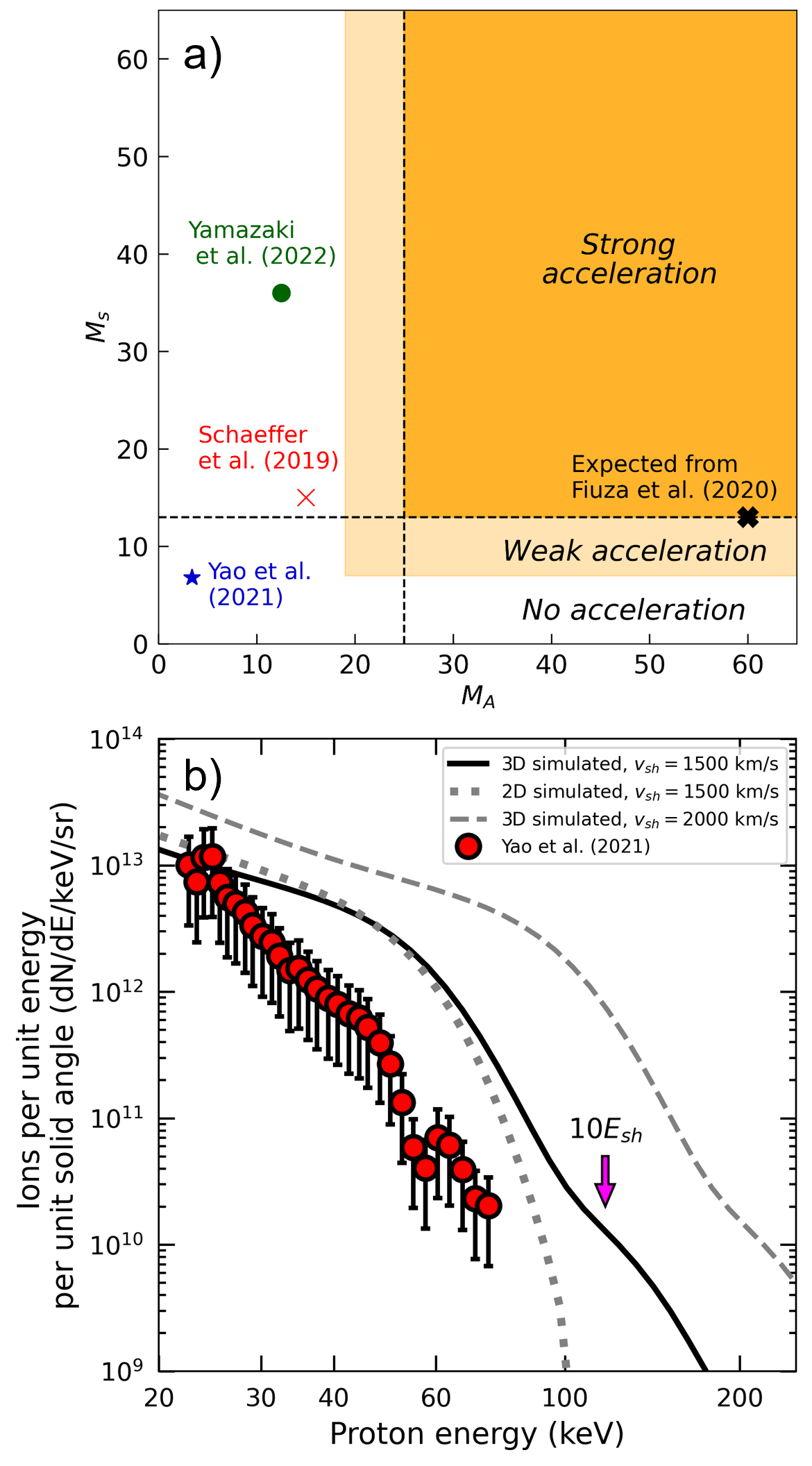}
    \caption{a) Phase diagram for 3D ion acceleration in $(M_A, M_s)$-space. Conditions covered in previous experiments by Schaeffer \textit{et al.}\cite{schaeffer+19} (2019), Yao \textit{et al.} (2021)\cite{yao+21}, and Yamazaki \textit{et al.} (2022)\cite{yamazaki+22} are shown, along with a realistic setup extrapolated from Fiuza \textit{et al.} (2020)\cite{fiuza+20}, assuming a pre-magnetized electron-proton plasma by a 10 T magnetic field and an upstream temperature of 60 eV; the $M_A$ and $M_s$ values obtained under these conditions, although not previously achieved experimentally, are representative of a plausible regime to be achieved at the NIF. b) Ion spectra measured in the laboratory\cite{yao+21} (red dots) at $(M_A=3.4, M_s=6.8)$ and predictions at $(M_A=25, M_s=13)$ based on our simulations (lines) with two shock velocities considered. The magenta arrow indicates the maximum energy $E_{\text{max}}=10E_{sh}$ for $v_{sh} = 1500$ km/s. Datasets reprinted with the authorization of the authors.
    } 
    \label{fig:experiments}
\end{figure}

Figure \ref{fig:experiments}a shows a phase diagram in $(M_A,M_s)$-space where we have identified that either strong, weak, or no acceleration occurs based on our simulations. This experimental setup closely resembles the conditions explored in the simulations presented here. However, we found that the system is not hypersonic nor long-lived enough to produce significant ion acceleration in 3-dimensions, and so previous simulations should be a good description of the acceleration process. Nevertheless, an experiment with a weaker magnetic field and longer time-frames may be able to access this regime.

\subsubsection{Yamazaki et al. at the Gekko-XII HIPER laser system}

The experiment conducted by Yamazaki et al. \cite{yamazaki+22} investigated the generation of quasi-perpendicular supercritical magnetized collisionless shocks using the Gekko-XII HIPER laser system. An aluminum target was irradiated with the laser, while the chamber was filled with nitrogen gas, which was subsequently ionized by photons emitted from the aluminum plasma, forming a magnetized plasma.  

An external magnetic field of $B_0 = 3.6$ T was applied, ensuring a nearly uniform field across the interaction region. The aluminum plasma expanded at an initial velocity of $v_{\mathrm{Al}} = 800$ km/s, compressing the nitrogen plasma and triggering the formation of a collisionless shock that propagated at a velocity of $v_{sh} = 400$ km/s.  

Shock conditions were sustained up to $t = 23$ ns after laser irradiation, revealing a well-defined shock foot and steep gradients characteristic of magnetized collisionless shocks. The derived shock parameters were $M_A \approx 12.5$ and $M_s \approx 36$, persisting for approximately $4 \omega_c^{-1}$. As shown in Figure \ref{fig:experiments}a, 1D and 2D simulations would capture the same physics as 3D ones.

\subsubsection{Yao et al. at the LULI2000 laser facility}

In the work reported by Yao et al.\cite{yao+21}, based on experiments conducted at the LULI2000 facility, a strong and uniform external magnetic field of 20 T was used to magnetize the ambient medium. The interaction medium consisted of hydrogen gas with an electron number density of $10^{18} \, \text{cm}^{-3}$. The shock front initially propagated at a velocity of approximately 1500 km/s, corresponding to $M_A = 3.4$ and $M_s = 6.8$ under the experimental conditions. This phase lasted for about 3 ns, equivalent to 6 $\omega_c^{-1}$, after which the shock velocity decreased to approximately 500 km/s.  

In this work, the authors investigated the accelerated ion spectra. During the shock phase, protons were accelerated to kinetic energies of up to 80 keV. The regime explored in this experiment remains within the range where 2D simulations provide a sufficient modeling framework for the underlying physical processes. Given the moderate values of $M_A$ and $M_s$ and the relatively short duration of the experiment, the production of ions with very high energies is not expected. 

This work provides an excellent point of comparison with our simulations. Figure \ref{fig:experiments}b compares ion spectra obtained experimentally, which was  measured after $\approx 5 \omega_c^{-1}$ with our calculations. We are interested in assessing if the accelerated particles we predict can be measured, at least in principle, with available instrumentation. Therefore, we will compare with one of our setups (not the most optimistic). We emphasize that we are not attempting a one-to-one comparison nor that we are accurately modeling the experiment discussed. 

To make the comparison, we assume the volume covered by the shock to be 2 mm$^3$ and place the spectrometer at a distance of 15 cm as stated in their report. Taking the spectra from the case ($M_A=25, \beta=5$), for two potential shock velocities, and after $10\omega_c^{-1}$, we find that the number of ions is similar to what is observed experimentally, but they show an altogether different spectrum. Also the 3D simulation shows a harder tail than the 2D one that should be above detectability. Additionally, we indicate the equivalent energy of $10 E_{sh}$ for a shock velocity of 1500 km/s. This implies that the instruments at the LULI2000 facility can be sensitive enough to detect the energetic ions.

\subsubsection{Fiuza et al. at the National Ignition Facility}

The work conducted by Fiuza et al. \cite{fiuza+20}, although not focused on ion dynamics in quasi-perpendicular shocks, serves as a valuable reference (in terms of characterstic plasma and shock conditions) for conditions that could be achieved at the National Ignition Facility (NIF). The experiments accelerated two identical counterstreaming plasma flows driven by 84 laser beams irradiating of two deuterated carbon (CD$_2$) targets. The plasma flows interacted in the central region, reaching velocities of $v_{sh} \approx 1800$ km/s. Non-thermal electrons were observed to be accelerated in the shock transition layer to energies reaching $\sim 500$ keV, exceeding the thermal energy by more than a factor of 100. Electron spectrometer measurements confirmed the presence of a power-law energy tail with a spectral index of $p \approx 3$.  

If these conditions were similar to a quasi-perpendicular shock experiment at the NIF, then for an electron-proton plasma premagnetized by a $10$ T magnetic field and a temperature of 60 eV, this setup could produce a shock at $(M_A \approx 60,M_s \approx 13)$ and persist for $\approx 24 \omega_c^{-1}$, which would satisfy the conditions for strong acceleration, as shown in Fig. \ref{fig:experiments}a. The values of $M_A$ and $M_s$ obtained under these conditions, although not previously achieved experimentally, are representative of characteristic parameters plausibly accessible at the NIF.


\subsubsection{Potential future experiments}
We propose a few possible parameter configurations for future experimental setups that should be equivalent to our Run B ($M_A=25$, $M_s=13$, $\beta=5$), assuming that it is possible to sustain the shock for 10 ns and achieve $N=10$ (so that $\tau_{exp}/N \omega_c^{-1} =1$) while considering different shock velocities. Under these conditions, the required magnetic field is 10 T for hydrogen and 20 T for other elements. Based on Figs. \ref{fig:scaling} and exploring different shock velocities, we find the following:  

\begin{itemize}  
    \item For $v_{sh}=500$ km/s, achieving the conditions of Run B is extremely challenging for any material due to the low shock velocity. This setup is not conducive to strong ion acceleration, as the required plasma parameters become impractical.  

    \item For $v_{sh}=1000$ km/s, using hydrogen as a target requires low temperature $T < 25$ eV, which is lower than previous experiments (e.g.\cite{Schaeffer2017}). Perhaps, a more feasible approach is to use a helium plasma, although full ionization requires a temperature of approximately 80 eV, which is higher than the upper limit. The required upstream electron density would be approximately $n_e=3.5 \times 10^{19}$ cm$^{-3}$.    

    \item For $v_{sh}=1500$ km/s, both hydrogen and helium setups become viable. If minimizing density is a priority, hydrogen is preferable, with $n_e \approx 1.5 \times 10^{19}$ cm$^{-3}$, provided that the temperature remains below 50 eV, a condition that has already been achieved experimentally \cite{yao+21}. If helium is used instead, higher temperatures of around 100 eV can be tolerated, producing a fully ionized medium with densities around $n_e=2.5 \times 10^{19}$ cm$^{-3}$.  
    \item For even higher velocities, such as $v_{sh}=2000$ km/s, both the applicable temperatures and densities shift accordingly. The required densities range from $n_e=0.8$ to $1.5 \times 10^{19}$ cm$^{-3}$, while temperatures vary between 70 and 200 eV.    
\end{itemize}  

If a higher $\tau_{exp}/N \omega_c^{-1}$ can be achieved, both $B_0$ and $n_e$ would decrease accordingly, as discussed in Sec. \ref{sec:scaling}. We note that these results can also be used as a guide to avoid having to deal with 3D simulations, which are currently prohibitively expensive to conduct with all the nuance a dedicated model needs. If this is the case, then future experiments can be planned to stay in the no acceleration phase of the $(M_A,M_s)$-space.

\section{Conclusions}
\label{sec:conclusions}
This work discusses the conditions necessary for ion acceleration in perpendicular magnetized collisionless shocks based on recent findings using 3D hybrid kinetic simulations, focusing in conditions relevant to laboratory experiments. By performing a parametric study using hybrid simulations, we identify thresholds of sonic and Alfv\'enic Mach numbers, together with relevant timescales, that dictate whether ion acceleration occurs.

We find that ion acceleration in perpendicular shocks requires a high Alfvénic ($M_A \gtrsim 25$) and hypersonic ($M_s \gtrsim 13$) Mach number that are equivalent to a moderately low plasma beta ($\beta \lesssim 5$). As demonstrated by the absence of a substantial non-thermal particle population in our simulations, significant ion acceleration does not occur if these thresholds are not satisfied. The presence of three-dimensional effects is essential for efficient acceleration, as they facilitate the scattering processes required for ions to re-cross the shock multiple times. However, for the experiments performed so far, 2D simulations remain sufficient to describe the main features of ion dynamics. 

We also explore the feasibility of recreating these conditions in laboratory settings, providing scaling relations that map astrophysical shock parameters to laser-driven plasma experiments. Our results indicate that existing facilities can potentially approach the strong acceleration regime, even considering the limitations in shock velocity and plasma magnetization. Experimental setups with high shock velocities ($v_{sh} \gtrsim 1000$ km/s) could be a promising setup, making it possible to observe efficient ion acceleration in controlled environments. Moreover, we have calculated particle spectra in experimentally-relevant conditions and found that these accelerated ions can be, at least in principle, detected with available instrumentation.

Future experiments could focus on optimizing plasma conditions to extend the shock evolution time and increase the number of ion gyro-periods captured. Additionally, by varying the ion composition, it may be possible to either enhance or suppress acceleration. This would provide a new method to test plasma astrophysics kinetic codes, connecting laboratory plasma physics with astrophysical shocks, and hopefully allowing to investigate CR acceleration mechanisms in controlled conditions.

\acknowledgments
We gratefully acknowledge D. Caprioli, J. Fuchs, D. B. Schaeffer and A. Spitkovsky for helpful discussions. 
Luca Orusa acknowledges the support of the Multimessenger Plasma Physics Center (MPPC), NSF grants PHY2206607. 

\section*{Authors' Statements}
Conflict of Interest Statement: 
The authors have no conflicts to disclose.

Data Availability Statement:
The data that support the findings of this study are available from the corresponding author upon reasonable request.

\bibliography{Total}

@article{Takabe2021,
author = {Takabe, Hideaki and Kuramitsu, Yasuhiro},
doi = {10.1017/hpl.2021.35},
issn = {20523289},
journal = {High Power Laser Science and Engineering},
title = {{Recent progress of laboratory astrophysics with intense lasers}},
volume = {9},
year = {2021}
}

@article{Germaschewski2016,
author = {Germaschewski, Kai and Fox, William and Abbott, Stephen and Ahmadi, Narges and Maynard, Kristofor and Wang, Liang and Ruhl, Hartmut and Bhattacharjee, Amitava},
doi = {10.1016/j.jcp.2016.05.013},
issn = {10902716},
journal = {Journal of Computational Physics},
pages = {305--326},
publisher = {Elsevier Inc.},
title = {{The Plasma Simulation Code: A modern particle-in-cell code with patch-based load-balancing}},
url = {http://dx.doi.org/10.1016/j.jcp.2016.05.013},
volume = {318},
year = {2016}
}

@article{Schaeffer2017,
author = {Schaeffer, D. B. and Fox, W. and Haberberger, D. and Fiksel, G. and Bhattacharjee, A. and Barnak, D. H. and Hu, S. X. and Germaschewski, K.},
doi = {10.1103/PhysRevLett.119.025001},
issn = {10797114},
journal = {Physical Review Letters},
number = {2},
pages = {1--6},
pmid = {28753335},
title = {{Generation and Evolution of High-Mach-Number Laser-Driven Magnetized Collisionless Shocks in the Laboratory}},
volume = {119},
year = {2017}
}

@article{Mcmillen2024,
author = {Mcmillen, K R and Heuer, P V and Gjevre, J M and Milder, A L and Charles, P and Filkins, T and Rinderknecht, H G and Froula, D H and Shaw, J L},
doi = {10.1063/5.0215756},
journal = {Review of Scientific Instruments},
pages = {073517},
publisher = {AIP Publishing, LLC},
title = {{Validation of predictive performance models for supersonic gas-jet nozzles at the Laboratory for Laser Energetics}},
url = {https://doi.org/10.1063/5.0215756},
volume = {95},
year = {2024}
}

@article{Barnak2018,
author = {Barnak, D. H. and Davies, J. R. and Fiksel, G. and Chang, P. Y. and Zabir, E. and Betti, R.},
doi = {10.1063/1.5012531},
issn = {10897623},
journal = {Review of Scientific Instruments},
number = {3},
pages = {2--7},
pmid = {29604743},
title = {{Increasing the magnetic-field capability of the magneto-inertial fusion electrical discharge system using an inductively coupled coil}},
url = {http://dx.doi.org/10.1063/1.5012531},
volume = {89},
year = {2018}
}

@article{yao+21,
   title={Laboratory evidence for proton energization by collisionless shock surfing},
   volume={17},
   ISSN={1745-2481},
   url={http://dx.doi.org/10.1038/s41567-021-01325-w},
   DOI={10.1038/s41567-021-01325-w},
   number={10},
   journal={Nature Physics},
   publisher={Springer Science and Business Media LLC},
   author={Yao, W. and Fazzini, A. and Chen, S. N. and Burdonov, K. and Antici, P. and Béard, J. and Bolaños, S. and Ciardi, A. and Diab, R. and Filippov, E. D. and Kisyov, S. and Lelasseux, V. and Miceli, M. and Moreno, Q. and Nastasa, V. and Orlando, S. and Pikuz, S. and Popescu, D. C. and Revet, G. and Ribeyre, X. and d’Humières, E. and Fuchs, J.},
   year={2021},
   month=aug, pages={1177–1182} }

@article{Chung2005,
author = {Chung, H. K. and Chen, M. H. and Morgan, W. L. and Ralchenko, Y. and Lee, R. W.},
doi = {10.1016/j.hedp.2005.07.001},
issn = {15741818},
journal = {High Energy Density Physics},
number = {1},
pages = {3--12},
title = {{FLYCHK: Generalized population kinetics and spectral model for rapid spectroscopic analysis for all elements}},
volume = {1},
year = {2005},
url = {https://www-nds.iaea.org/exfor/servlet/X4sZvd?file=https://www-amdis.iaea.org/FLYCHK/ZBAR/zp006.zvd}
}

@article{Schaeffer2012,
author = {Schaeffer, D. B. and Everson, E. T. and Winske, D. and Constantin, C. G. and Bondarenko, A. S. and Morton, L. A. and Flippo, K. A. and Montgomery, D. S. and Gaillard, S. A. and Niemann, C.},
doi = {10.1063/1.4736846},
journal = {Physics of Plasmas},
number = {7},
title = {{Generation of magnetized collisionless shocks by a novel, laser-driven magnetic piston}},
volume = {19},
year = {2012}
}

@article{hada+03,
author = {Hada, Tohru and Oonishi, Makiko and Lembège, Bertrand and Savoini, Philippe},
title = {Shock front nonstationarity of supercritical perpendicular shocks},
journal = {Journal of Geophysical Research: Space Physics},
volume = {108},
number = {A6},
pages = {},
doi = {https://doi.org/10.1029/2002JA009339},
url = {https://agupubs.onlinelibrary.wiley.com/doi/abs/10.1029/2002JA009339},
eprint = {https://agupubs.onlinelibrary.wiley.com/doi/pdf/10.1029/2002JA009339},
year = {2003}
}

@article{bell78a,
	Author = {{Bell}, A.~R.},
	Journal = {MNRAS},
	Month = jan,
	Pages = {147-156},
	Title = {{The acceleration of cosmic rays in shock fronts. I}},
	Url = {https://ui.adsabs.harvard.edu/abs/1978MNRAS.182..147B/abstract},
	Volume = 182,
	Year = 1978}

@article{blandford+78,
	Author = {{Blandford}, R.~D. and {Ostriker}, J.~P.},
	Doi = {10.1086/182658},
	Journal = {ApJL},
	Month = apr,
	Pages = {L29-L32},
	Title = {{Particle acceleration by astrophysical shocks}},
	Url = {https://ui.adsabs.harvard.edu/abs/1978ApJ...221L..29B},
	Volume = {221},
	Year = {1978}}

@article{caprioli+14a,
	Adsurl = {http://adsabs.harvard.edu/abs/2014ApJ...783...91C},
	Archiveprefix = {arXiv},
	Author = {{Caprioli}, D. and {Spitkovsky}, A.},
	Doi = {10.1088/0004-637X/783/2/91},
	Eid = {91},
	Eprint = {1310.2943},
	Journal = {\apj},
	Month = mar,
	Pages = {91},
	Primaryclass = {astro-ph.HE},
	Title = {{Simulations of Ion Acceleration at Non-relativistic Shocks: I. Acceleration Efficiency}},
	Volume = {783},
	Year = {2014}}

@article{caprioli+14b,
	Adsurl = {http://adsabs.harvard.edu/abs/2014ApJ...794...46C},
	Archiveprefix = {arXiv},
	Author = {{Caprioli}, D. and {Spitkovsky}, A.},
	Doi = {10.1088/0004-637X/794/1/46},
	Eid = {46},
	Eprint = {1401.7679},
	Journal = {\apj},
	Month = oct,
	Pages = {46},
	Primaryclass = {astro-ph.HE},
	Title = {{Simulations of Ion Acceleration at Non-relativistic Shocks: II. Magnetic Field Amplification}},
	Volume = {794},
	Year = {2014}}

@article{caprioli+14c,
	Adsurl = {http://adsabs.harvard.edu/abs/2014ApJ...794...47C},
	Archiveprefix = {arXiv},
	Author = {{Caprioli}, D. and {Spitkovsky}, A.},
	Doi = {10.1088/0004-637X/794/1/47},
	Eid = {47},
	Eprint = {1407.2261},
	Journal = {\apj},
	Month = oct,
	Pages = {47},
	Primaryclass = {astro-ph.HE},
	Title = {{Simulations of Ion Acceleration at Non-relativistic Shocks. III. Particle Diffusion}},
	Volume = {794},
	Year = {2014}}

@article{caprioli+15,
	Archiveprefix = {arXiv},
	Author = {{Caprioli}, D. and {Pop}, A.R. and {Spitkovsky}, A.},
	Eid = {28},
	Eprint = {1409.8291},
	Journal = {\apjl},
	Month = jan,
	Pages = {28},
	Primaryclass = {astro-ph.HE},
	Title = {{Simulations and Theory of Ion Injection at Non-relativistic Collisionless Shocks}},
	Volume = {798},
	Year = {2015}}

@article{kucharek+91,
	Adsurl = {http://adsabs.harvard.edu/abs/1991JGR....9621195K},
	Author = {{Kucharek}, H. and {Scholer}, M.},
	Doi = {10.1029/91JA02321},
	Journal = {\jgr},
	Month = dec,
	Pages = {21195},
	Title = {{Origin of diffuse superthermal ions at quasi-parallel supercritical collisionless shocks}},
	Volume = 96,
	Year = 1991}

@article{amano+07,
	Adsurl = {http://adsabs.harvard.edu/abs/2007ApJ...661..190A},
	Author = {{Amano}, T. and {Hoshino}, M.},
	Doi = {10.1086/513599},
	Eprint = {arXiv:astro-ph/0612204},
	Journal = {\apj},
	Month = may,
	Pages = {190-202},
	Title = {{Electron Injection at High Mach Number Quasi-perpendicular Shocks: Surfing and Drift Acceleration}},
	Volume = 661,
	Year = 2007}

@article{gamil+08,
	Author = {{Cassam-Chena{\"\i}}, G. and {Hughes}, J.~P. and {Reynoso}, E.~M. and {Badenes}, C. and {Moffett}, D.},
	Doi = {10.1086/588015},
	Eprint = {arXiv:0803.0805},
	Journal = {\apj},
	Month = jun,
	Pages = {1180-1197},
	Title = {{Morphological Evidence for Azimuthal Variations of the Cosmic-Ray Ion Acceleration at the Blast Wave of SN 1006}},
	Url = {http://cdsads.u-strasbg.fr/abs/2008ApJ...680.1180C},
	Volume = 680,
	Year = 2008}

@article{schaeffer+19,
  title = {Direct Observations of Particle Dynamics in Magnetized Collisionless Shock Precursors in Laser-Produced Plasmas},
  author = {Schaeffer, D. B. and Fox, W. and Follett, R. K. and Fiksel, G. and Li, C. K. and Matteucci, J. and Bhattacharjee, A. and Germaschewski, K.},
  journal = {Phys. Rev. Lett.},
  volume = {122},
  issue = {24},
  pages = {245001},
  numpages = {6},
  year = {2019},
  month = {Jun},
  publisher = {American Physical Society},
  doi = {10.1103/PhysRevLett.122.245001},
  url = {https://link.aps.org/doi/10.1103/PhysRevLett.122.245001}
}

@article{yamazaki+22,
    author = {{Yamazaki}, R. and {Matsukiyo}, S. and {Morita}, T. and {Tanaka}, S.~J. and {Umeda}, T. and {Aihara}, K. and {Edamoto}, M. and {Egashira}, S. and {Hatsuyama}, R. and {Higuchi}, T. and others},
    title = "{High-power laser experiment forming a supercritical collisionless shock in a magnetized uniform plasma at rest}",
    eprint = "2201.07976",
    archivePrefix = "arXiv",
    primaryClass = "physics.plasm-ph",
    doi = "10.1103/PhysRevE.105.025203",
    journal = "Phys. Rev. E",
    volume = "105",
    number = "2",
    pages = "025203",
    year = "2022"
}

@article{caprioli12,
	Adsurl = {http://adsabs.harvard.edu/abs/2012JCAP...07..038C},
	Archiveprefix = {arXiv},
	Author = {{Caprioli}, D.},
	Doi = {10.1088/1475-7516/2012/07/038},
	Eid = {038},
	Eprint = {1206.1360},
	Journal = {\jcap},
	Month = jul,
	Pages = {38},
	Primaryclass = {astro-ph.HE},
	Title = {{Cosmic-ray acceleration in supernova remnants: non-linear theory revised}},
	Volume = {7},
	Year = {2012}}

@article{giacalone+93,
	Adsurl = {http://adsabs.harvard.edu/abs/1993ApJ...402..550G},
	Author = {{Giacalone}, J. and {Burgess}, D. and {Schwartz}, S.~J. and {Ellison}, D.~C.},
	Doi = {10.1086/172157},
	Journal = {\apj},
	Month = jan,
	Pages = {550-559},
	Title = {{Ion injection and acceleration at parallel shocks - Comparisons of self-consistent plasma simulations with existing theories}},
	Volume = 402,
	Year = 1993}

@article{giacalone+97,
	Adsurl = {http://adsabs.harvard.edu/abs/1997JGR...10219789G},
	Author = {{Giacalone}, J. and {Burgess}, D. and {Schwartz}, S.~J. and {Ellison}, D.~C. and {Bennett}, L.},
	Doi = {10.1029/97JA01529},
	Journal = {\jgr},
	Month = sep,
	Pages = {19789-19804},
	Title = {{Injection and acceleration of thermal protons at quasi-parallel shocks: A hybrid simulation parameter survey}},
	Volume = 102,
	Year = 1997}

@article{gargate+07,
	    author = {{Gargat{\'e}}, L. and {Bingham}, R. and {Fonseca}, R.~A. and {Silva}, L.~O.},
        title = "{dHybrid: A massively parallel code for hybrid simulations of space plasmas}",
      journal = {Computer Physics Communications},
         year = 2007,
        month = mar,
       volume = {176},
       number = {6},
        pages = {419-425},
          doi = {10.1016/j.cpc.2006.11.013},
archivePrefix = {arXiv},
       eprint = {physics/0611174},
 primaryClass = {physics.plasm-ph},
       adsurl = {https://ui.adsabs.harvard.edu/abs/2007CoPhC.176..419G}
}

@article{rothenflug+04,
	Author = {{Rothenflug}, R. and {Ballet}, J. and {Dubner}, G. and {Giacani}, E. and {Decourchelle}, A. and {Ferrando}, P.},
	Doi = {10.1051/0004-6361:20047104},
	Journal = {A\&A},
	Month = oct,
	Pages = {121-131},
	Title = {{Geometry of the non-thermal emission in SN 1006. Azimuthal variations of cosmic-ray acceleration}},
	Url = {http://adsabs.harvard.edu/abs/2004A26A...425..121R},
	Volume = 425,
	Year = 2004}

@inproceedings{axford+77p,
	Author = {{Axford}, W.~I. and {Leer}, E. and {Skadron}, G.},
	Booktitle = {\emph{Acceleration of Cosmic Rays at Shock Fronts}},
	Pages = {273-+},
	Series = {International Cosmic Ray Conference},
	Title = {{Acceleration of Cosmic Rays at Shock Fronts (Abstract)}},
	Url = {http://adsabs.harvard.edu/abs/1977ICRC....2..273A},
	Volume = 2,
	Year = 1977}

@article{marcowith+16,
	Adsurl = {https://ui.adsabs.harvard.edu/abs/2016RPPh...79d6901M},
	Archiveprefix = {arXiv},
	Author = {{Marcowith}, A. and {Bret}, A. and {Bykov}, A. and {Dieckman}, M.~E. and {O'C Drury}, L. and {Lemb{\`e}ge}, B. and {Lemoine}, M. and {Morlino}, G. and {Murphy}, G. and {Pelletier}, G. and {Plotnikov}, I. and {Reville}, B. and {Riquelme}, M. and {Sironi}, L. and {Stockem Novo}, A.},
	Doi = {10.1088/0034-4885/79/4/046901},
	Eid = {046901},
	Eprint = {1604.00318},
	Journal = {Reports on Progress in Physics},
	Month = apr,
	Number = {4},
	Pages = {046901},
	Primaryclass = {astro-ph.HE},
	Title = {{The microphysics of collisionless shock waves}},
	Volume = {79},
	Year = 2016}

@article{SN1006HESS,
	Adsurl = {http://adsabs.harvard.edu/abs/2010A%26A...516A..62A},
	Archiveprefix = {arXiv},
	author={Acero, Fabio and Aharonian, F and Akhperjanian, AG and Anton, G and De Almeida, U Barres and Bazer-Bachi, AR and Becherini, Yvonne and Behera, B and Beilicke, M and Bernl{\"o}hr, K and others},
	Doi = {10.1051/0004-6361/200913916},
	Eprint = {1004.2124},
	Journal = {A\&A},
	Month = jun,
	Pages = {A62+},
	Primaryclass = {astro-ph.HE},
	Title = {{First detection of VHE {$\gamma$}-rays from SN 1006 by HESS}},
	Volume = 516,
	Year = 2010}

@article{morlino+12,
	Adsurl = {http://adsabs.harvard.edu/abs/2012A%26A...538A..81M},
	Archiveprefix = {arXiv},
	Author = {{Morlino}, G. and {Caprioli}, D.},
	Doi = {10.1051/0004-6361/201117855},
	Eid = {A81},
	Eprint = {arXiv:1105.6342},
	Journal = {A\&A},
	Month = feb,
	Pages = {A81},
	Primaryclass = {astro-ph.HE},
	Title = {{Strong evidence for hadron acceleration in Tycho's supernova remnant}},
	Volume = {538},
	Year = {2012}}

@article{matsumoto+15,
	Adsurl = {http://adsabs.harvard.edu/abs/2015Sci...347..974M},
	Author = {{Matsumoto}, Y. and {Amano}, T. and {Kato}, T.~N. and {Hoshino}, M.},
	Doi = {10.1126/science.1260168},
	Journal = {Science},
	Month = feb,
	Pages = {974-978},
	Title = {{Stochastic electron acceleration during spontaneous turbulent reconnection in a strong shock wave}},
	Volume = {347},
	Year = {2015}}

@article{caprioli+17,
	Adsurl = {https://ui.adsabs.harvard.edu/abs/2017PhRvL.119q1101C},
	Archiveprefix = {arXiv},
	Author = {{Caprioli}, Damiano and {Yi}, Dennis T. and {Spitkovsky}, Anatoly},
	Doi = {10.1103/PhysRevLett.119.171101},
	Eid = {171101},
	Eprint = {1704.08252},
	Journal = {\prl},
	Month = {8},
	Number = {17},
	Pages = {171101},
	Primaryclass = {astro-ph.HE},
	Title = {{Chemical Enhancements in Shock-Accelerated Particles: Ab initio Simulations}},
	Volume = {119},
	Year = {2017}}

@article{brunetti+14,
	Archiveprefix = {arXiv},
	Author = {Brunetti, G. and Jones, T. W.},
	Doi = {10.1142/S0218271814300079},
	Eid = {1430007-98},
	Eprint = {1401.7519},
	Journal = {International Journal of Modern Physics D},
	Month = mar,
	Pages = {1430007-98},
	Title = {Cosmic Rays in Galaxy Clusters and Their Nonthermal Emission},
	Volume = {23},
	Year = {2014}}

@article{caprioli+18,
	Archiveprefix = {arXiv},
	Author = {Caprioli, D. and Zhang, H. and Spitkovsky, A.},
	Eprint = {1801.01510},
	Journal = {\jpp},
	Month = jan,
	Primaryclass = {astro-ph.HE},
	Title = {Diffusive Shock Re-Acceleration},
	Url = {http://adsabs.harvard.edu/abs/2018arXiv180101510C},
	Year = {2018}}

@article{giacalone05,
	Author = {Giacalone, J.},
	Doi = {10.1086/432510},
	Journal = {\apjl},
	Month = jul,
	Pages = {L37-L40},
	Title = {The Efficient Acceleration of Thermal Protons by Perpendicular Shocks},
	Url = {http://adsabs.harvard.edu/abs/2005ApJ...628L..37G},
	Volume = {628},
	Year = {2005}}

@article{chevalier+06,
	Author = {Chevalier, R. A. and Fransson, C.},
	Doi = {10.1086/507606},
	Eprint = {astro-ph/0607196},
	Journal = {\apj},
	Month = nov,
	Pages = {381-391},
	Title = {Circumstellar Emission from Type Ib and Ic Supernovae},
	Url = {http://adsabs.harvard.edu/abs/2006ApJ...651..381C},
	Volume = {651},
	Year = {2006}}

@article{guo+14a,
	Archiveprefix = {arXiv},
	Author = {Guo, X. and Sironi, L. and Narayan, R.},
	Doi = {10.1088/0004-637X/794/2/153},
	Eid = {153},
	Eprint = {1406.5190},
	Journal = {\apj},
	Month = oct,
	Pages = {153},
	Primaryclass = {astro-ph.HE},
	Title = {Non-thermal Electron Acceleration in Low Mach Number Collisionless Shocks. I. Particle Energy Spectra and Acceleration Mechanism},
	Url = {http://adsabs.harvard.edu/abs/2014ApJ...794..153G},
	Volume = {794},
	Year = {2014}}

@article{guo+14b,
	Archiveprefix = {arXiv},
	Author = {Guo, X. and Sironi, L. and Narayan, R.},
	Doi = {10.1088/0004-637X/797/1/47},
	Eid = {47},
	Eprint = {1409.7393},
	Journal = {\apj},
	Month = dec,
	Pages = {47},
	Primaryclass = {astro-ph.HE},
	Title = {Non-thermal Electron Acceleration in Low Mach Number Collisionless Shocks. II. Firehose-mediated Fermi Acceleration and its Dependence on Pre-shock Conditions},
	Url = {http://adsabs.harvard.edu/abs/2014ApJ...797...47G},
	Volume = {797},
	Year = {2014}}

@article{haggerty+19a,
	Adsurl = {https://ui.adsabs.harvard.edu/abs/2019ApJ...887..165H},
	Archiveprefix = {arXiv},
	Author = {{Haggerty}, Colby C. and {Caprioli}, Damiano},
	Doi = {10.3847/1538-4357/ab58c8},
	Eid = {165},
	Eprint = {1909.05255},
	Journal = {\apj},
	Month = {12},
	Number = {2},
	Pages = {165},
	Primaryclass = {astro-ph.HE},
	Title = {{dHybridR: A Hybrid Particle-in-cell Code Including Relativistic Ion Dynamics}},
	Volume = {887},
	Year = {2019}}

@article{xu+20,
	Adsurl = {https://ui.adsabs.harvard.edu/abs/2020ApJ...897L..41X},
	Archiveprefix = {arXiv},
	Author = {{Xu}, Rui and {Spitkovsky}, Anatoly and {Caprioli}, Damiano},
	Doi = {10.3847/2041-8213/aba11e},
	Eid = {L41},
	Eprint = {1908.07890},
	Journal = {\apjl},
	Month = jul,
	Number = {2},
	Pages = {L41},
	Primaryclass = {astro-ph.HE},
	Title = {{Electron Acceleration in One-dimensional Nonrelativistic Quasi-perpendicular Collisionless Shocks}},
	Volume = {897},
	Year = 2020}

@article{johlander+21,
       author = {{Johlander}, A. and {Battarbee}, M. and {Vaivads}, A. and {Turc}, L. and {Pfau-Kempf}, Y. and {Ganse}, U. and {Grandin}, M. and {Dubart}, M. and {Khotyaintsev}, Yu. V. and {Caprioli}, D. and {Haggerty}, C. and {Schwartz}, S.~J. and {Giles}, B.~L. and {Palmroth}, M.},
        title = "{Ion Acceleration Efficiency at the Earth's Bow Shock: Observations and Simulation Results}",
      journal = {\apj},
         year = 2021,
        month = jun,
       volume = {914},
       number = {2},
          eid = {82},
        pages = {82},
          doi = {10.3847/1538-4357/abfafc},
       adsurl = {https://ui.adsabs.harvard.edu/abs/2021ApJ...914...82J}
}

@article{haggerty+20,
	Adsurl = {https://ui.adsabs.harvard.edu/abs/2020ApJ...905....1H},
	Archiveprefix = {arXiv},
	Author = {{Haggerty}, Colby C. and {Caprioli}, Damiano},
	Doi = {10.3847/1538-4357/abbe06},
	Eid = {1},
	Eprint = {2008.12308},
	Journal = {\apj},
	Month = dec,
	Number = {1},
	Pages = {1},
	Primaryclass = {astro-ph.HE},
	Title = {{Kinetic Simulations of Cosmic-Ray-modified Shocks. I. Hydrodynamics}},
	Volume = {905},
	Year = 2020}

@article{caprioli+20,
	Adsurl = {https://ui.adsabs.harvard.edu/abs/2020ApJ...905....2C},
	Archiveprefix = {arXiv},
	Author = {{Caprioli}, Damiano and {Haggerty}, Colby C. and {Blasi}, Pasquale},
	Doi = {10.3847/1538-4357/abbe05},
	Eid = {2},
	Eprint = {2009.00007},
	Journal = {\apj},
	Month = dec,
	Number = {1},
	Pages = {2},
	Primaryclass = {astro-ph.HE},
	Title = {{Kinetic Simulations of Cosmic-Ray-modified Shocks. II. Particle Spectra}},
	Volume = {905},
	Year = 2020}

@article{bohdan+19a,
	Adsurl = {https://ui.adsabs.harvard.edu/abs/2019ApJ...878....5B},
	Archiveprefix = {arXiv},
	Author = {{Bohdan}, Artem and {Niemiec}, Jacek and {Pohl}, Martin and {Matsumoto}, Yosuke and {Amano}, Takanobu and {Hoshino}, Masahiro},
	Doi = {10.3847/1538-4357/ab1b6d},
	Eid = {5},
	Eprint = {1904.13153},
	Journal = {\apj},
	Month = jun,
	Number = {1},
	Pages = {5},
	Primaryclass = {astro-ph.HE},
	Title = {{Kinetic Simulations of Nonrelativistic Perpendicular Shocks of Young Supernova Remnants. I. Electron Shock-surfing Acceleration}},
	Volume = {878},
	Year = 2019}

@article{boehly+95,
    author = {Boehly, T. R. and Craxton, R. S. and Hinterman, T. H. and Kelly, J. H. and Kessler, T. J. and Kumpan, S. A. and Letzring, S. A. and McCrory, R. L. and Morse, S. F. B. and Seka, W. and Skupsky, S. and Soures, J. M. and Verdon, C. P.},
    title = {The upgrade to the OMEGA laser system},
    journal = {Review of Scientific Instruments},
    volume = {66},
    number = {1},
    pages = {508-510},
    year = {1995},
    month = {01},
    issn = {0034-6748},
    doi = {10.1063/1.1146333},
    url = {https://doi.org/10.1063/1.1146333},
    eprint = {https://pubs.aip.org/aip/rsi/article-pdf/66/1/508/19254304/508\_1\_online.pdf},
}

@ARTICLE{lembege+04,
       author = {{Lembege}, B. and {Giacalone}, J. and {Scholer}, M. and {Hada}, T. and {Hoshino}, M. and {Krasnoselskikh}, V. and {Kucharek}, H. and {Savoini}, P. and {Terasawa}, T.},
        title = "{Selected Problems in Collisionless-Shock Physics}",
      journal = {\ssr},
         year = 2004,
        month = jan,
       volume = {110},
       number = {3},
        pages = {161-226},
          doi = {10.1023/B:SPAC.0000023372.12232.b7},
       adsurl = {https://ui.adsabs.harvard.edu/abs/2004SSRv..110..161L}
}

@ARTICLE{vanweeren+10,
       author = {{van Weeren}, Reinout J. and {R{\"o}ttgering}, Huub J.~A. and {Br{\"u}ggen}, Marcus and {Hoeft}, Matthias},
        title = "{Particle Acceleration on Megaparsec Scales in a Merging Galaxy Cluster}",
      journal = {Science},
         year = 2010,
        month = oct,
       volume = {330},
       number = {6002},
        pages = {347},
          doi = {10.1126/science.1194293},
archivePrefix = {arXiv},
       eprint = {1010.4306},
 primaryClass = {astro-ph.CO},
       adsurl = {https://ui.adsabs.harvard.edu/abs/2010Sci...330..347V}
}

@ARTICLE{bocchino+11,
       author = {{Bocchino}, F. and {Orlando}, S. and {Miceli}, M. and {Petruk}, O.},
        title = "{Constraints on the local interstellar magnetic field from non-thermal emission of SN1006}",
      journal = {\aap},
         year = 2011,
        month = jul,
       volume = {531},
          eid = {A129},
        pages = {A129},
          doi = {10.1051/0004-6361/201016341},
archivePrefix = {arXiv},
       eprint = {1105.2689},
 primaryClass = {astro-ph.HE},
       adsurl = {https://ui.adsabs.harvard.edu/abs/2011A&A...531A.129B}
}

@article{govoni+04,
	doi = {10.1086/382674},
	url = {https://doi.org/10.1086/382674},
	year = 2004,
	month = {apr},
	publisher = {American Astronomical Society},
	volume = {605},
	number = {2},
	pages = {695--708},
	author = {F. Govoni and M. Markevitch and A. Vikhlinin and L. VanSpeybroeck and L. Feretti and G. Giovannini},
	title = {{ChandraTemperature} Maps for Galaxy Clusters with Radio Halos},
	journal = {The Astrophysical Journal}
}

@ARTICLE{giuffrida+22,
       author = {{Giuffrida}, Roberta and {Miceli}, Marco and {Caprioli}, Damiano and {Decourchelle}, Anne and {Vink}, Jacco and {Orlando}, Salvatore and {Bocchino}, Fabrizio and {Greco}, Emanuele and {Peres}, Giovanni},
        title = "{The supernova remnant SN 1006 as a Galactic particle accelerator}",
      journal = {Nat. Comm.},
         year = 2022,
        month = aug,
       volume = {13},
          eid = {5098},
        pages = {5098},
          doi = {10.1038/s41467-022-32781-4},
archivePrefix = {arXiv},
       eprint = {2208.14491},
 primaryClass = {astro-ph.HE},
       adsurl = {https://ui.adsabs.harvard.edu/abs/2022NatCo..13.5098G}
}

@ARTICLE{fiuza+20,
       author = {{Fiuza}, F. and {Swadling}, G.~F. and {Grassi}, A. and {Rinderknecht}, H.~G. and {Higginson}, D.~P. and {Ryutov}, D.~D. and {Bruulsema}, C. and {Drake}, R.~P. and {Funk}, S. and {Glenzer}, S. and {Gregori}, G. and {Li}, C.~K. and {Pollock}, B.~B. and {Remington}, B.~A. and {Ross}, J.~S. and {Rozmus}, W. and {Sakawa}, Y. and {Spitkovsky}, A. and {Wilks}, S. and {Park}, H. -S.},
        title = "{Electron acceleration in laboratory-produced turbulent collisionless shocks}",
      journal = {Nature Physics},
         year = 2020,
        month = jun,
       volume = {16},
       number = {9},
        pages = {916-920},
          doi = {10.1038/s41567-020-0919-4},
       adsurl = {https://ui.adsabs.harvard.edu/abs/2020NatPh..16..916F}
}

@article{orusa+25,
   title={The Role of Three-dimensional Effects on Ion Injection and Acceleration in Perpendicular Shocks},
   volume={1001},
   ISSN={1538-4357},
   url={http://dx.doi.org/10.3847/1538-4357/ae563e},
   DOI={10.3847/1538-4357/ae563e},
   number={2},
   journal={The Astrophysical Journal},
   publisher={American Astronomical Society},
   author={Orusa, Luca and Caprioli, Damiano and Sironi, Lorenzo and Spitkovsky, Anatoly},
   year={2026},
   month=Apr, pages={158} }

@article{bohdan+21,
  title = {Magnetic Field Amplification by the Weibel Instability at Planetary and Astrophysical Shocks with High Mach Number},
  author = {Bohdan, Artem and Pohl, Martin and Niemiec, Jacek and Morris, Paul J. and Matsumoto, Yosuke and Amano, Takanobu and Hoshino, Masahiro and Sulaiman, Ali},
  journal = {Phys. Rev. Lett.},
  volume = {126},
  issue = {9},
  pages = {095101},
  numpages = {6},
  year = {2021},
  month = {Mar},
  publisher = {American Physical Society},
  doi = {10.1103/PhysRevLett.126.095101},
  url = {https://link.aps.org/doi/10.1103/PhysRevLett.126.095101}
}

@article{shimada+00,
doi = {10.1086/318161},
url = {https://dx.doi.org/10.1086/318161},
year = {2000},
month = {nov},
publisher = {},
volume = {543},
number = {1},
pages = {L67},
author = {N. Shimada and M. Hoshino},
title = {Strong Electron Acceleration at High Mach Number Shock
Waves: Simulation Study of Electron
Dynamics},
journal = {The Astrophysical Journal},
}

@article{kato+10,
	doi = {10.1088/0004-637x/721/1/828},
	url = {https://doi.org/10.1088\%2F0004-637x\%2F721\%2F1\%2F828},
	year = 2010,
	month = {sep},
	publisher = {American Astronomical Society},
	volume = {721},
	number = {1},
	pages = {828--842},
	author = {Tsunehiko N. Kato and Hideaki Takabe},
	title = {Nonrelativistic collisionless shocks in weakly magnetized electron--ion plasmas: two-dimensional particle-in-cell simulation of perpendicular shock
},
	journal = {The Astrophysical Journal}
}

@article{matsumoto+17,
	doi = {10.1103/physrevlett.119.105101},
	url = {https://doi.org/10.1103\%2Fphysrevlett.119.105101},
	year = 2017,
	month = {sep},
	publisher = {American Physical Society ({APS})},
	volume = {119},
	number = {10},
	author = {Yosuke Matsumoto and Takanobu Amano and Tsunehiko N. Kato and Masahiro Hoshino},
	title = {Electron Surfing and Drift Accelerations in a Weibel-Dominated High-Mach-Number Shock},
	journal = {Physical Review Letters}
}

@article{lalti+22,
author = {Lalti, A. and Khotyaintsev, Yu. V. and Dimmock, A. P. and Johlander, A. and Graham, D. B. and Olshevsky, V.},
title = {A Database of MMS Bow Shock Crossings Compiled Using Machine Learning},
journal = {Journal of Geophysical Research: Space Physics},
volume = {127},
number = {8},
pages = {e2022JA030454},
doi = {https://doi.org/10.1029/2022JA030454},
year = {2022}
}

@ARTICLE{orusa+23,
       author = {{Orusa}, Luca and {Caprioli}, Damiano},
        title = "{Fast Particle Acceleration in 3D Hybrid Simulations of Quasiperpendicular Shocks}",
      journal = {\prl},
         year = 2023,
        month = sep,
       volume = {131},
       number = {9},
          eid = {095201},
        pages = {095201},
          doi = {10.1103/PhysRevLett.131.095201},
archivePrefix = {arXiv},
       eprint = {2305.10511},
 primaryClass = {astro-ph.HE},
       adsurl = {https://ui.adsabs.harvard.edu/abs/2023PhRvL.131i5201O}
}

@ARTICLE{morris+23,
       author = {{Morris}, Paul J. and {Bohdan}, Artem and {Weidl}, Martin S. and {Tsirou}, Michelle and {Fulat}, Karol and {Pohl}, Martin},
        title = "{Pre-acceleration in the Electron Foreshock. II. Oblique Whistler Waves}",
      journal = {\apj},
         year = 2023,
        month = feb,
       volume = {944},
       number = {1},
          eid = {13},
        pages = {13},
          doi = {10.3847/1538-4357/acaec8},
archivePrefix = {arXiv},
       eprint = {2301.00872},
 primaryClass = {physics.plasm-ph},
       adsurl = {https://ui.adsabs.harvard.edu/abs/2023ApJ...944...13M}
}

@ARTICLE{wilson+16b,
   author = {{Wilson III}, L.~B. and {Sibeck}, D.~G. and {Turner}, D.~L. and
	{Osmane}, A. and {Caprioli}, D. and {Angelopoulos}, V.},
    title = "{Relativistic electrons produced by foreshock disturbances observed upstream of the Earth's bow shock}",
  journal = {Phys. Rev. Lett.},
     year = 2016,
    month = nov,
   volume = 117,
   number = 21,
      eid = {215101},
    pages = {215101},
      doi = {10.1103/PhysRevLett.117.215101},
     note = {Editors' Suggestion},
   adsurl = {https://ui.adsabs.harvard.edu/abs/2016PhRvL.117u5101W}
}

@ARTICLE{ha+23,
       author = {{Ha}, Ji-Hoon and {Ryu}, Dongsu and {Kang}, Hyesung},
        title = "{Cosmic-Ray Acceleration and Nonthermal Radiation at Accretion Shocks in the Outer Regions of Galaxy Clusters}",
      journal = {\apj},
         year = 2023,
        month = feb,
       volume = {943},
       number = {2},
          eid = {119},
        pages = {119},
          doi = {10.3847/1538-4357/acabbe},
archivePrefix = {arXiv},
       eprint = {2210.16817},
 primaryClass = {astro-ph.HE},
       adsurl = {https://ui.adsabs.harvard.edu/abs/2023ApJ...943..119H}
}

@ARTICLE{ha+22,
       author = {{Ha}, Ji-Hoon and {Ryu}, Dongsu and {Kang}, Hyesung and {Kim}, Sunjung},
        title = "{Electron Preacceleration at Weak Quasi-perpendicular Intracluster Shocks: Effects of Preexisting Nonthermal Electrons}",
      journal = {\apj},
         year = 2022,
        month = jan,
       volume = {925},
       number = {1},
          eid = {88},
        pages = {88},
          doi = {10.3847/1538-4357/ac3bc0},
archivePrefix = {arXiv},
       eprint = {2110.14236},
 primaryClass = {astro-ph.HE},
       adsurl = {https://ui.adsabs.harvard.edu/abs/2022ApJ...925...88H}
}

@ARTICLE{ha+21,
       author = {{Ha}, Ji-Hoon and {Kim}, Sunjung and {Ryu}, Dongsu and {Kang}, Hyesung},
        title = "{Effects of Multiscale Plasma Waves on Electron Preacceleration at Weak Quasi-perpendicular Intracluster Shocks}",
      journal = {\apj},
         year = 2021,
        month = jul,
       volume = {915},
       number = {1},
          eid = {18},
        pages = {18},
          doi = {10.3847/1538-4357/abfb68},
archivePrefix = {arXiv},
       eprint = {2102.03042},
 primaryClass = {astro-ph.HE},
       adsurl = {https://ui.adsabs.harvard.edu/abs/2021ApJ...915...18H}
}

@ARTICLE{kang+19,
       author = {{Kang}, Hyesung and {Ryu}, Dongsu and {Ha}, Ji-Hoon},
        title = "{Electron Preacceleration in Weak Quasi-perpendicular Shocks in High-beta Intracluster Medium}",
      journal = {\apj},
         year = 2019,
        month = may,
       volume = {876},
       number = {1},
          eid = {79},
        pages = {79},
          doi = {10.3847/1538-4357/ab16d1},
archivePrefix = {arXiv},
       eprint = {1901.04173},
 primaryClass = {astro-ph.HE},
       adsurl = {https://ui.adsabs.harvard.edu/abs/2019ApJ...876...79K}
}

@ARTICLE{amano+22,
       author = {{Amano}, Takanobu and {Matsumoto}, Yosuke and {Bohdan}, Artem and {Kobzar}, Oleh and {Matsukiyo}, Shuichi and {Oka}, Mitsuo and {Niemiec}, Jacek and {Pohl}, Martin and {Hoshino}, Masahiro},
        title = "{Nonthermal electron acceleration at collisionless quasi-perpendicular shocks}",
      journal = {Reviews of Modern Plasma Physics},
         year = 2022,
        month = dec,
       volume = {6},
       number = {1},
          eid = {29},
        pages = {29},
          doi = {10.1007/s41614-022-00093-1},
archivePrefix = {arXiv},
       eprint = {2209.03521},
 primaryClass = {astro-ph.HE},
       adsurl = {https://ui.adsabs.harvard.edu/abs/2022RvMPP...6...29A}
}

@ARTICLE{amano+09a,
       author = {{Amano}, Takanobu and {Hoshino}, Masahiro},
        title = "{Electron Shock Surfing Acceleration in Multidimensions: Two-Dimensional Particle-in-Cell Simulation of Collisionless Perpendicular Shock}",
      journal = {\apj},
         year = 2009,
        month = jan,
       volume = {690},
       number = {1},
        pages = {244-251},
          doi = {10.1088/0004-637X/690/1/244},
archivePrefix = {arXiv},
       eprint = {0805.1098},
 primaryClass = {astro-ph},
       adsurl = {https://ui.adsabs.harvard.edu/abs/2009ApJ...690..244A}
}

@ARTICLE{jones+98,
       author = {{Jones}, Frank C. and {Jokipii}, J. Randy and {Baring}, Matthew G.},
        title = "{Charged-Particle Motion in Electromagnetic Fields Having at Least One Ignorable Spatial Coordinate}",
      journal = {\apj},
         year = 1998,
        month = dec,
       volume = {509},
       number = {1},
        pages = {238-243},
          doi = {10.1086/306480},
archivePrefix = {arXiv},
       eprint = {astro-ph/9808103},
 primaryClass = {astro-ph},
       adsurl = {https://ui.adsabs.harvard.edu/abs/1998ApJ...509..238J}
}

@ARTICLE{kumar+21,
       author = {{Kumar}, Naveen and {Reville}, Brian},
        title = "{Nonthermal Particle Acceleration at Highly Oblique Nonrelativistic Shocks}",
      journal = {\apjl},
         year = 2021,
        month = nov,
       volume = {921},
       number = {1},
          eid = {L14},
        pages = {L14},
          doi = {10.3847/2041-8213/ac30e0},
archivePrefix = {arXiv},
       eprint = {2110.09939},
 primaryClass = {physics.plasm-ph},
       adsurl = {https://ui.adsabs.harvard.edu/abs/2021ApJ...921L..14K}
}

@ARTICLE{boula+24,
       author = {{Boula}, S.~S. and {Niemiec}, J. and {Amano}, T. and {Kobzar}, O.},
        title = "{Quasi-perpendicular shocks of galaxy clusters in hybrid kinetic simulations. The structure of the shocks}",
      journal = {\aap},
         year = 2024,
        month = apr,
       volume = {684},
          eid = {A129},
        pages = {A129},
          doi = {10.1051/0004-6361/202349091},
archivePrefix = {arXiv},
       eprint = {2402.00571},
 primaryClass = {astro-ph.HE},
       adsurl = {https://ui.adsabs.harvard.edu/abs/2024A&A...684A.129B}
}

@article{fujita+01,
   title={Nonthermal Emission from Accreting and Merging Clusters of Galaxies},
   volume={563},
   ISSN={1538-4357},
   url={http://dx.doi.org/10.1086/324030},
   DOI={10.1086/324030},
   number={2},
   journal={The Astrophysical Journal},
   publisher={American Astronomical Society},
   author={Fujita, Yutaka and Sarazin, Craig L.},
   year={2001},
   month=dec, pages={660–672} }

@ARTICLE{willson70,
       author = {{Willson}, M.~A.~G.},
        title = "{Radio observations of the cluster of galaxies in Coma Berenices - the 5C4 survey.}",
      journal = {\mnras},
         year = 1970,
        month = jan,
       volume = {151},
        pages = {1},
          doi = {10.1093/mnras/151.1.1},
       adsurl = {https://ui.adsabs.harvard.edu/abs/1970MNRAS.151....1W}
}

@article{lindner+14,
   title={THE RADIO RELICS AND HALO OF EL GORDO, A MASSIVEz= 0.870 CLUSTER MERGER},
   volume={786},
   ISSN={1538-4357},
   url={http://dx.doi.org/10.1088/0004-637X/786/1/49},
   DOI={10.1088/0004-637x/786/1/49},
   number={1},
   journal={The Astrophysical Journal},
   publisher={American Astronomical Society},
   author={Lindner, Robert R. and Baker, Andrew J. and Hughes, John P. and Battaglia, Nick and Gupta, Neeraj and Knowles, Kenda and Marriage, Tobias A. and Menanteau, Felipe and Moodley, Kavilan and Reese, Erik D. and Srianand, Raghunathan},
   year={2014},
   month=apr, pages={49} }

@BOOK{tidman+71,
       author = {{Tidman}, D.~A. and {Krall}, N.~A.},
        title = "{Shock waves in collisionless plasmas}",
         year = 1971,
       adsurl = {https://ui.adsabs.harvard.edu/abs/1971swcp.book.....T}
}

@ARTICLE{jikei+24,
       author = {{Jikei}, Taiki and {Amano}, Takanobu and {Matsumoto}, Yosuke},
        title = "{Enhanced Magnetic Field Amplification by Ion-beam Weibel Instability in Weakly Magnetized Astrophysical Shocks}",
      journal = {\apj},
         year = 2024,
        month = feb,
       volume = {961},
       number = {2},
          eid = {157},
        pages = {157},
          doi = {10.3847/1538-4357/ad1594},
archivePrefix = {arXiv},
       eprint = {2312.07933},
 primaryClass = {astro-ph.HE},
       adsurl = {https://ui.adsabs.harvard.edu/abs/2024ApJ...961..157J}
}

@ARTICLE{nishigai+21,
       author = {{Nishigai}, Takuro and {Amano}, Takanobu},
        title = "{Mach number dependence of ion-scale kinetic instability at collisionless perpendicular shock: Condition for Weibel-dominated shock}",
      journal = {Physics of Plasmas},
         year = 2021,
        month = jul,
       volume = {28},
       number = {7},
          eid = {072903},
        pages = {072903},
          doi = {10.1063/5.0051269},
archivePrefix = {arXiv},
       eprint = {2107.02404},
 primaryClass = {physics.plasm-ph},
       adsurl = {https://ui.adsabs.harvard.edu/abs/2021PhPl...28g2903N}
}

@PREAMBLE{
 "\providecommand{\noopsort}[1]{}" 
 # "\providecommand{\singleletter}[1]{#1}%" 
}
\end{document}